\documentclass[12pt,letterpaper]{JHEP3}

\usepackage{amsmath}
\usepackage{cancel}
\usepackage{epsfig}
\usepackage[vcentermath]{syoungtab}

\newcommand{\tensor}{\otimes}

\newcommand{\ayng}[2]{\overbrace{\yng(#2,#2)\raise 3.2pt\hbox{\,$\cdots$\yng(1)}}^{#1}}
\newcommand{\byng}[1]{\overbrace{\yng(1)\,\cdots\yng(1)}^{#1}}

\newcommand{\V}[2]{{\cal V}^{#1}_{\Yboxdim5.7pt\styng(#2)}}
\newcommand{\Vd}[3]{{\cal V}^{#3}_{(#1)[#2]}}
\newcommand{\Vdb}[3]{{\overline{\cal V}}^{#3}_{(#1)[#2]}}
\newcommand{\cv}{{\cal V}}
\newcommand{\ocv}{{\overline{\cal V}}}

\preprint{Brown-HET-1577}

\title{\Large The Morphology of ${\cal{N}} = 6$ Chern-Simons Theory}

\author{Georgios Papathanasiou and Marcus Spradlin\\
Department of Physics\\
Brown University, Providence, Rhode Island 02912, USA\\

E-mail: \email{Georgios\_Papathanasiou@brown.edu},
\email{Marcus\_Spradlin@brown.edu}}

\abstract{
We tabulate various properties
of the `language' of ${\cal{N}} = 6$
Chern-Simons Theory, in the sense of Polyakov.  Specifically
we enumerate and compute character formulas for all syllables of
up to four letters, i.e.~all irreducible representations of
$OSp(6|4)$ built from up to four fundamental fields of
the ABJM theory.
We also present all tensor product decompositions for up
to four singletons and list the
(cyclically invariant) four-letter words,
which correspond to
single-trace operators of length four.
As an application of these results we
use the two-loop dilatation operator to
compute the leading correction
to the Hagedorn temperature of the weakly-coupled planar ABJM theory
on ${\mathbb{R}} \times S^2$.
}

\begin{document}

\section{Introduction}

It is difficult to overstate the importance of the role that
maximally superconformal field theories have played in deepening
our understanding of string and field theories, and the relations between
them.  To date the vast majority of work has focused on four-dimensional
gauge theories describing the worldvolume of D3-branes, but much
progress has recently been made on three-dimensional theories
describing the worldvolume of M2-branes.  This progress has been
made possible by the discovery~\cite{Bagger:2007jr}
(see also~\cite{Gustavsson:2007vu})
of highly supersymmetric conformal
theories in three dimensions,
extending earlier
attempts~\cite{Schwarz:2004yj} with superconformal Chern-Simons theories.

This paper is concerned with the conformal ${\cal{N}} = 6$ supersymmetric
Chern-Simons matter theory constructed by ABJM~\cite{Aharony:2008ug}
(see also~\cite{Benna:2008zy}, and~\cite{Gaiotto:2007qi} for
related earlier work).
The ABJM theory is a three-dimensional $U(N) \times U(N)$ gauge
theory with four complex scalars and their fermionic
partners in the
bi-fundamental representation and
gauge fields with Chern-Simons levels $+k$ and $-k$.
The theory has a 't Hooft limit in which $N,k \to \infty$
with $\lambda=N/k$ fixed, similar to the story in ${\cal{N}} = 4$
Yang-Mills theory (SYM).

Indeed the strong similarity to SYM has allowed many
tools, such as the language of spin chains and
methods from
integrability (such
as~\cite{Minahan:2002ve,Beisert:2003yb,Arutyunov:2004vx,Beisert:2006ez}) which have been so successful
in exploring the structure of planar SYM, to be applied to the ABJM
theory as well.
In particular the anomalous dimensions of local operators
in the theory are apparently encoded in an integrable
spin chain
Hamiltonian (see~\cite{Minahan:2008hf,Gaiotto:2008cg,Bak:2008cp,Bak:2008vd,
Spill:2008yr,Kristjansen:2008ib,Ahn:2009zg,Gromov:2009tv}
for related recent work),
and an exact magnon S-matrix for this spin
chain has been proposed in~\cite{Gromov:2008qe} (see also~\cite{Ahn:2008aa}).
In several respects however the story of integrability in the ABJM theory
is slightly more complicated
than that in SYM, one of which (see also~\cite{Krishnan:2008zs}) is
the fact that anomalous dimensions first show up at two loops.
The full two-loop dilatation operator,
which has recently been constructed in~\cite{Zwiebel:2009vb,Minahan:2009te},
has both nearest-neighbor and next-to-nearest neighbor interactions,
making it necessarily more complicated than the nearest-neighbor
one-loop dilatation operator~\cite{Beisert:2003jj} of SYM.

There is a very beautiful and useful analogy~\cite{Polyakov:2001af}
between the counting of local gauge-invariant operators
in gauge theories such as SYM and linguistics.
The elementary fields (and their derivatives) are thought of as
`letters' which are strung together inside single trace operators
as `words', products of which can then be thought of as
`sentences'.
Since the ABJM elementary fields transform in either the
$(N,\overline{N})$ or the $(\overline{N},N)$ bi-fundamental representations
they must appear in alternating order inside any
single-trace operator.
If we wish to extend the linguistic analogy to this case we could
perhaps say that the ABJM alphabet is divided into
consonants and vowels, comprising the respectively
the two $OSp(6|4)$ singleton representations (which are conjugate
to each other).  Every word in the ABJM language has even length
and consists of alternating vowels and consonants.

One of the purposes of this paper is to lay some groundwork
for detailed spectroscopic analysis of the ABJM theory through
two loops using the results of~\cite{Zwiebel:2009vb,Minahan:2009te}.
To this end we first review in section~\ref{section_oscillator_construction}
the oscillator construction for $OSp(6|4)$.  Then
in section~\ref{sectionthree} we enumerate
all syllables of up to four letters (i.e.~all irreducible representations
of $OSp(6|4)$ built from up to four generations of superoscillators)
and also calculate their characters
(these may also be found in the exhaustive reference~\cite{Dolan:2008vc}).

In section~\ref{tensorsection}
we present all tensor product decompositions for products of up
to four singletons.
These tensor product results are useful for analysis of the two-loop
dilatation operator since its building block, the
``Hamiltonian density'' $D_2$, is an operator which acts
simultaneously on three
adjacent sites of the spin chain, alternately occupied
by the two singletons $\cv^1$ and $\ocv^1$.  Due to $OSp(6|4)$ symmetry
the Hamiltonian
density can be written in block-diagonal
form with no mixing between irreducible representations of different
quantum numbers in the tensor product decomposition
of $\cv^1 \tensor \ocv^1 \tensor \cv^1$.
For the four-site tensor products we also calculate the
irreducible representations in the tensor product decomposition
of $(\cv^1 \tensor \ocv^1)^2$ which are symmetric under interchange
of the two $\cv^1 \tensor \ocv^1$ factors.
This representation content corresponds to the physical spectrum
of gauge invariant operators of length four in the ABJM theory---i.e.~they
are the four-letter words in the ABJM language.
For completeness we include section~\ref{subsectorsection} in
which all of the abovementioned results are tabulated for the
$OSp(4|2)$ subsector.

We defer more detailed spectroscopy for later
work, only mentioning it here as one motivation for this work.
However in section~\ref{peeksection}
we present a concrete result which follows
rather easily from the explicit form of the
two-loop dilatation operator and the information presented in this paper:
a calculation of the two-loop correction to the Hagedorn temperature
$T_{\rm H}$ of the planar ABJM theory on $S^2$, with the result
$\delta \log T_{\rm H} = 2 \lambda^2 ( \sqrt{2} - 1)$.
Our results might also be useful for studying higher spin symmetry
in the free ABJM theory, following
for example~\cite{Bianchi:2003wx,Beisert:2004di}.

\section{Oscillator Construction for $OSp(6|4)$}
\label{section_oscillator_construction}

In this section we review the oscillator construction for the $OSp(6|4)$
supergroup
due to Gunaydin and Hyun~\cite{Gunaydin:1988kz},
together with the particular
notations which are convenient for our purposes\footnote{An alternative
approach, using the method of Kac,
may be found in~\cite{Morel:1984hr}.}.
We begin with the bosonic $Sp(4,\mathbb{R})$ and $SO(6)$
subgroups before building up to the full $OSp(6|4)$.

\subsection{$Sp(4,\mathbb{R})\simeq SO(3,2)$}

The $Sp(4,\mathbb{R})\simeq SO(3,2)$ generators can be expressed in terms
of a set of $f=2p+\epsilon$ ($\epsilon=0,1$)
``generations'' of bosonic annihilation operators
$a_i(r), b_i(r), c_i$, $(i=1,2, r=1,2,...,p)$ and their hermitian conjugate
creation operators
$a^i(r)=a_i^\dagger(r), b^i(r)=b_i^\dagger(r), c^i=c_i^\dagger$,
transforming respectively
covariantly and contravariantly under $U(2)$.
The number of $a$ and $b$ oscillators $p$ can be
any integer greater or equal to zero, whereas we only have either
zero or one $c$ oscillators, according to the value of $\epsilon$.
The oscillators obey the usual commutation commutation relations, the
only nonvanishing ones being
\begin{equation}
\left[ a_i(r) , a^j(s) \right] = \delta_i^j \delta_{rs} \,,
\qquad
\left[ b_i(r) , b^j(s) \right] = \delta_i^j \delta_{rs} \,,
\qquad
\left[ c_i , c^j \right]       = \delta_i^j \,.
\end{equation}
The $Sp(4,\mathbb{R})$ generators are then given as
\begin{equation}
\begin{aligned}
P^{ij}     &= \vec{a}^i \cdot \vec{b}^j + \vec{a}^j \cdot \vec{b}^i + \epsilon \; c^i c^j
&&=  \yng(2) \,,\\
K_{ij}     &= \vec{a}_i \cdot \vec{b}_j + \vec{a}_j \cdot \vec{b}_i + \epsilon \; c_i c_j
&&=  (P^{ji})^\dagger \,,\\
I{}^i{}_j  &= \vec{a}^i \cdot \vec{a}_j + \vec{b}_j \cdot \vec{b}^i + \frac{\epsilon}{2} \left( c^i c_j + c_j c^i \right)
&&= (I{}^j{}_i)^\dagger \,,
\end{aligned}
\label{Sp4Rgenerators}
\end{equation}
where we have adopted the vector notation
$\vec a_i=\left( a_i(1),a_i(2),...,a_i(p)\right)$ implying
$\vec{a}_i \cdot \vec{b}_j=\sum_{r=1}^p a_i(r) b_j(r)$ and so on.
The association of $P^{ij}$ with the Young tableau (YT) $\yng(2)$ is
included here for convenience and will be explained shortly.

In this particular basis the $Sp(4,\mathbb{R})$ algebra is
\begin{equation}
\begin{aligned}
\left[ K_{ij} , P^{kl} \right]
&= \delta_j^l I{}^k{}_i + \delta_i^k I{}^l{}_j + \delta_j^k I{}^l{}_i + \delta_i^l I{}^k{}_j \,,\\
\left[ I{}^i{}_j , P^{kl} \right]
&= \delta_j^k P^{il} + \delta_j^l P^{ik} \,,\\
\left[ I{}^i{}_j , K_{kl} \right]
&= -\delta_k^i K_{jl}-\delta_l^i K_{jk} \,,\\
\left[ I{}^i{}_j , I{}^k{}_l \right]
&= \delta_j^k I{}^i{}_l - \delta_l^i I{}^k{}_j \,,
\end{aligned}
\label{Sp4Ralgebra}
\end{equation}
and we can immediately recognize the $I{}^i{}_j$ as generators of the
maximal compact subgroup $U(2) \subset Sp(4,\mathbb{R})$.
In terms of the bosonic number operators which we define as
\begin{equation}
\begin{aligned}
N_{B_i} &= \vec{a}^i \cdot \vec{a}_i + \vec{b}^i \cdot \vec{b}_i + \epsilon \; c^i c_i
& \text{(no sum on $i$)} \,,\\
N_B &= N_{B_1}+N_{B_2} \,,
\end{aligned}
\label{boson_number_op}
\end{equation}
the diagonal entries of the $U(2)$ generators may be rewritten as
\begin{equation}
I{}^1{}_1 = N_{B_1} + \frac{1}{2} f \,, \qquad
I{}^2{}_2 = N_{B_2} + \frac{1}{2} f \,.
\end{equation}

The form of the commutators~(\ref{Sp4Ralgebra}) is essentially the same
as that of the three-dimensional conformal group in the spinor basis (see for
example~\cite{Dolan:2008vc}) with the identification of
\begin{equation}
\Delta = \frac{1}{2} I{}^i{}_i = \frac{1}{2}(N_B + f)
\label{Delta}
\end{equation}
as the
dilatation operator\footnote{The observant reader may worry that $\Delta$
is a compact generator, and that in general we construct states that have
definite $U(2)\simeq SO(2)\times SO(3)$ charges instead of
$SO(1,1)\times SO(1,2)$ ones.
However it has been proven in~\cite{Fabbri:1999ay} that there exists
a rotation which maps the eigenstates of one subgroup to those
of the other while preserving their eigenvalues,
similarly to what happens for the four-dimensional
conformal group which had been proven earlier in~\cite{Gunaydin:1998jc}.}
and $L^i_j=I^i_j-\delta^i_j\,\Delta$, $P^{ij}$, and $K_{ij}$ as
the generators of rotations, translations, and special conformal
transformations respectively.

We are interested in representations of $Sp(4,\mathbb{R})$ for
which the spectrum of $\Delta$ is
bounded from below. Each such representation can be characterized
by its lowest-weight state (LWS) $|\Omega\rangle$, a state within the
multiplet that
\begin{enumerate}
\item{is annihilated by all $K_{ij}$, and}
\item{transforms irreducibly under the $U(2)$ subgroup.}
\end{enumerate}
Of course the ``vacuum'' $|0\rangle$, defined as usual
as the state which is annihilated by all lowering operators
$a_i(r), b_i(r), c_i$ is a suitable LWS, but not the only one. The first
condition implies that there exist additional lowest-weight
states which can be expressed as
linear combinations of raising operators acting on $|0\rangle$. Since the
raising operators transform in the fundamental two-dimensional representation
of $U(2)$, the irreducible representations arising from combinations of
$M$ raising operators are related to representations of the permutation
group $S_M$, and the second condition implies that we simply have to
symmetrize and antisymmetrize the combinations of raising operators
appropriately in order to get a LWS.

In this manner the $U(2)$ YT description of a LWS arises
naturally in conjunction with the actual symmetrization and antisymmetrization
of raising operators, and only differs from the proper $SU(2)$ YT
in that we do not need to discard two-box columns,
as they provide information for
the $U(1)$ charge which corresponds to the scaling dimension $\Delta$. If $(m_1,m_2)$ denote the number of boxes in the (first,second) rows
of a $U(2)$ YT, then the scaling dimension $\Delta$ and the
$SU(2)$ spin $j$ of the corresponding LWS are given by
\begin{equation}
(\Delta,j) = (\frac{1}{2} (m_1+m_2+f),\frac{1}{2}(m_1-m_2)) \,.
\label{Sp4_labels}
\end{equation}
As an example, it can be shown that for $f=2$ the only inequivalent
$Sp(4,\mathbb{R})$ lowest-weight
states are of the form
\begin{equation}
\begin{aligned}
a^{i_1} a^{i_2} \cdots a^{i_k} |0\rangle &= \overbrace{\yng(2)\cdots\yng(1)}^k &\qquad (\Delta,j) &=(1+\frac{k}{2},\frac{k}{2}) &\qquad k=0,1,2,\ldots \,,\\
(a^i b^j - a^j b^i) |0\rangle &= \yng(1,1) & (\Delta,j) &= (2,0) \,,
\end{aligned}
\end{equation}
where we have also indicated the corresponding Young tableaux and the
Cartan charges $(\Delta,j)$ of the LWS.

Given the LWS $|\Omega\rangle$
of any representation, a basis $R$ for the entire
representation is generated by acting on it with the various $P^{ij}$,
\begin{equation}
R=\{|\Omega\rangle,P^{ij}|\Omega\rangle,P^{lm}P^{ij}|\Omega\rangle,\ldots\} \,.
\end{equation}
There is no restriction on how many times we can act with $P^{ij}$, hence we
produce an an infinite-dimensional representation (a
manifestation of the noncompact nature of $Sp(4,\mathbb{R})$).
If we are interested in
the $U(2)$ content of each of the basis vectors, this can be determined by
considering symmetric tensor products of $P^{ij}$, due to the
fact that the $P^{ij}$ commute with each other.
Using the identification of $P^{ij}$ with the YT $\yng(2)$ as indicated
above in~(\ref{Sp4Rgenerators}) we have for example
\begin{equation}
P^{ij} = \yng(2) \,,
\quad
(P^{ij})^{ 2}_+ = \yng(2,2)+\yng(4) \,,
\quad
(P^{ij})^{3}_+ = \yng(4,2)+\yng(6) \,,
\quad
\text{etc,}
\end{equation}
where the subscript $+$ indicates that only the totally
symmetric representations in the tensor product decomposition contribute.
In this manner one can obtain
the full $U(2)$ content of the multiplet by tensoring arbitrarily
high symmetric
powers of $\yng(2)$ with the YT of the LWS, following the usual
decomposition rules.

\subsection{$SO(6)\simeq SU(4)$}

The oscillator construction we will be describing here is a particular
realization for $n=3$ of the general method for $SO(2n)$ groups, however
since the fundamental fields of the ABJM theory transform in the complex
{\bf 4} and $\bf \bar 4$ representations rather than the real {\bf 6}
representation, the language of the locally isomorphic $SU(4)$ will be
more suitable for the presentation.

This time the building blocks for the generators will be fermionic
oscillators. We again have $f=2p+\epsilon$, ($\epsilon=0,1$) annihilation
operators
$\alpha_\mu(r), \beta_\mu(r), \gamma_\mu$, $(\mu=1,2,3,\, r=1,2,...,p)$
and their hermitian conjugate creation operators
$\alpha^\mu(r)=\alpha_\mu^\dagger(r),
\beta^\mu(r)=\beta_\mu^\dagger(r),
\gamma^\mu=\gamma_\mu^\dagger$ transforming in the conjugate fundamental and
fundamental of $U(3)$ respectively. The only nonvanishing
anticommutation relations are
\begin{equation}
\left\{ \alpha_\mu(r) , \alpha^\nu(s) \right\} = \delta_\mu^\nu \delta_{rs} \,,
\qquad
\left\{ \beta_\mu(r) , \beta^\nu(s) \right\} = \delta_\mu^\nu \delta_{rs} \,,
\qquad
\left\{ \gamma_\mu , \gamma^\nu \right\} = \delta_\mu^\nu \,.
\end{equation}
With respect to these oscillators the $SU(4)$ generators are given by
\begin{equation}
\begin{aligned}
A^{\mu\nu}    &= \vec{\alpha}^\mu \cdot \vec{\beta}^\nu - \vec{\alpha}^\nu \cdot \vec{\beta}^\mu + \epsilon \; \gamma^\mu \gamma^\nu
&&=  \yng(1,1) \,,\\
A_{\mu\nu}    &= \vec{\alpha}_\mu \cdot \vec{\beta}_\nu - \vec{\alpha}_\nu \cdot \vec{\beta}_\mu + \epsilon \; \gamma_\mu \gamma_\nu
&&=  (A^{\nu\mu})^\dagger \,,\\
U{}^\mu{}_\nu &= \vec{\alpha}^\mu \cdot \vec{\alpha}_\nu - \vec{\beta}_\nu \cdot \vec{\beta}^\mu + \frac{\epsilon}{2} \left( \gamma^\mu \gamma_\nu - \gamma_\nu \gamma^\mu \right) & & =  (U{}^\nu{}_\mu)^\dagger \,,
\end{aligned}
\label{SO6generators}
\end{equation}
from which the corresponding algebra in this particular basis follows:
\begin{equation}
\begin{aligned}
\left[ A_{\mu\nu} , A^{\rho\sigma} \right]
&= -\delta_\mu^\sigma U{}^\rho{}_\nu + \delta_\mu^\rho U{}^\sigma{}_\nu - \delta_\nu^\rho U{}^\sigma{}_\mu +\delta_\nu^\sigma U{}^\rho{}_\mu \,,\\
\left[ U{}^\mu{}_\nu,A^{\rho \sigma} \right]
&= \delta^\rho_\nu A^{\mu \sigma}+\delta^\sigma_\nu A^{\rho\mu} \,,\\
\left[ U{}^\mu{}_\nu, A_{\rho\sigma} \right]
&= -\delta^\mu_\rho A_{\nu \sigma}-\delta_\sigma^\mu A_{\rho\nu} \,,\\
\left[ U{}^\mu{}_\nu , U{}^\rho{}_\sigma \right]
&= \delta_\nu^\rho U{}^\mu{}_\sigma - \delta_\sigma^\mu U{}^\rho{}_\nu \,.
\end{aligned}
\label{SO6algebra}
\end{equation}
It is evident
evident from the last
line that the $U{}^\mu{}_\nu$ are generators of a $U(3)$ subgroup.

The relations~(\ref{SO6algebra}) are the so-called split form of
the commutation relations because one $U(3)$ is singled out.
They can be recast into the standard $SU(4)$ form by defining
\begin{equation}
\begin{aligned}
R{}^\mu{}_\nu &= U{}^\mu{}_\nu - \frac{1}{2} \delta^\mu_\nu U{}^\lambda{}_\lambda \,,
& R{}^\mu{}_4 &= +\frac{1}{2}\, \epsilon^{\mu\rho\sigma} A{}_{\rho\sigma}
\quad \Rightarrow \quad
A{}_{\mu\nu} = \epsilon_{\mu\nu\rho} R{}^\rho{}_4 \,,\\
R{}^4{}_4 &= \frac{1}{2} U{}^\lambda{}_\lambda \,,
& R{}^4{}_\mu &= -\frac{1}{2}\, \epsilon_{\mu\rho\sigma} A{}^{\rho\sigma}
\quad \Rightarrow \quad
A{}^{\mu\nu} = \epsilon^{\mu\nu\rho} R{}^4{}_\rho \,,\\
\end{aligned}
\end{equation}
which as a consequence of~(\ref{SO6algebra}) obey
\begin{equation}
\left[ R{}^\alpha{}_\beta , R{}^\gamma{}_\delta \right] =
\delta_\beta^\gamma R{}^\alpha{}_\delta - \delta_\delta^\alpha R{}^\gamma{}_\beta \,,
\qquad
\text{where here
$\alpha, \beta, \gamma, \delta=1,\ldots,4$} \,,
\end{equation}
with $R{}^\alpha{}_\alpha=0$, $(R{}^\beta{}_\alpha)^\dagger=R{}^\alpha{}_\beta$
as required for $SU(4)$.

For future use we mention
here the forms of the diagonal $U$ and $R$ generators in
terms of the fermionic number operators
\begin{equation}
\begin{aligned}
N_{F_\mu} &= \vec{\alpha}^\mu \cdot \vec{\alpha}_\mu + \vec{\beta}^\mu \cdot \vec{\beta}_\mu + \epsilon \; \gamma^\mu\gamma_\mu
& \text{(no sum on $\mu$)} \,,\\
N_F &= N_{F_1} + N_{F_2} + N_{F_3}
\end{aligned}
\end{equation}
which are
\begin{equation}
\begin{aligned}
U{}^\mu{}_\mu &= N_{F_\mu} - \frac{1}{2} f & \text{(no sum on $\mu$)} \,,\\
R{}^\mu{}_\mu &= N_{F_\mu} - N_F + \frac{1}{4} f & \text{(no sum on $\mu$)} \,,\\
R{}^4{}_4 &= N_F - \frac{3}{4} f\,.
\end{aligned}
\end{equation}

As the oscillator method for the construction of representations is very
general and applies to a variety of groups and supergroups, see for
example~\cite{Gunaydin:1988kz,Fernando:2004jt}, the essential
features of the $SU(4)$ case are similar to what we saw for the
$Sp(4,\mathbb{R})$ representations in the previous section. In a nutshell,
after we define the ``vacuum'' $|0\rangle$ as the state annihilated
by all lowering operators $\alpha_\mu(r), \beta_\mu(r), \gamma_\mu$, we
look for lowest-weight states $|\Omega\rangle$ that are
annihilated by $A_{\mu\nu}$ and transform irreducibly under $U(3)$. These are
expressed in terms of properly symmetrized and antisymmetrized
creation operators, described naturally in terms of $U(3)$ Young tableaux,
whose only difference from the proper $SU(3)$ Young tableaux is that we no
longer discard three-box columns.

Then by acting repeatedly on a LWS with various $A^{\mu\nu}$ we build a basis
for an irreducible representation of $SU(4)$, and the $U(3)$ content of
each of the representation may be found by tensoring the symmetric
powers of
$A^{\mu\nu}$,
\begin{equation}
A{}^{\mu\nu} = \yng(1,1) \,,
\qquad
(A{}^{\mu\nu})^{ 2}_+ =\yng(2,2) \,,
\qquad
(A{}^{\mu\nu})^{ 3}_+ = \yng(3,3) \,,
\end{equation}
and so on, with the YT of the LWS.

The only major difference compared
to the $Sp(4,\mathbb{R})$ case is that since we
have $3f$ fermionic oscillators, $(A^{\mu\nu})^k=0$ for $k>\frac{3}{2} f$
and hence the representations will now be finite-dimensional, reflecting the
compactness of $SU(4)$. Therefore each representation has a
highest-weight state (HWS)
which is annihilated by all $A^{\mu\nu}$,
transforms irreducibly under $U(3)$, and is related to the LWS
by unitarity. We will make use of this relation, as the labels of a
representation are related to the weights of its HWS. In particular, we
will use $SU(4)$ Dynkin labels to characterize
representations\footnote{$SU(4)$ and $SO(6)$ have the same Dynkin
diagram with just the ordering of the first two roots switched, which
translates into the relation
$[d_1,d_2,d_3]^{SU(4)}=[d_2,d_1,d_3]^{SO(6)}$ for
their Dynkin labels.}, and as we'll see later on
these are related to the labels $(l_1,l_2,l_3)$ denoting the number of
boxes in the first, second and third rows of the $U(3)$ LWS YT by
\begin{equation}
[d_1,d_2,d_3]=[f-l_1-l_2,l_2-l_3,l_1-l_2] \,.
\label{SU4_yt2dynk}
\end{equation}
We should mention that in the literature another labeling convention is
also widely used (for example in~\cite{Bhattacharya:2008bja,Dolan:2008vc})
which is based on
eigenvalues under rotations in three orthogonal planes in ${\mathbb{R}}^6$
(hence more suited to the $SO(6)$ description of the algebra), known as
$SO(6)$ Gelfand-Zetlin labels $(r_1,r_2,r_3)$ . A basic property they
obey is that $r_1\ge r_2\ge|r_3|$, and in our conventions they are related
to the $SU(4)$ Dynkin labels by
\begin{equation}
(r_1,r_2,r_3)=(d_2+\frac{1}{2}(d_3+d_1),\frac{1}{2}(d_3+d_1),\frac{1}{2}(d_3-d_1)) \,.
\end{equation}
We illustrate the basic steps of the procedure described above with an
example. For $f=1$ the only possible lowest-weight
states and Young tableaux
are\footnote{We use ``1'' to denote the singlet of the $U(3)$ subgroup,
not the singlet of the full group.}
\begin{equation}
\begin{aligned}
|0\rangle &= 1 &\qquad [d_1,d_2,d_3] &=[1,0,0] &\qquad (r_1,r_2,r_3) &=(\frac{1}{2},\frac{1}{2},-\frac{1}{2}) \,,\\
\gamma^\mu|0\rangle &= \yng(1) & \qquad [d_1,d_2,d_3] &=[0,0,1] & (r_1,r_2,r_3) &= (\frac{1}{2},\frac{1}{2},\frac{1}{2}) \,,
\end{aligned}
\end{equation}
where we have also indicated both the Dynkin and Gelfand-Zetlin labels
of the corresponding representations.

\subsection{$OSp(6|4)$ and Super-Young Tableaux}
\label{sytsection}

For the full $OSp(6|4)$ superalgebra one needs to combine the
$Sp(4,\mathbb{R})$ and $SO(6)$ oscillators described in the
previous sections into
$U(2|3)$ contravariant and covariant superoscillators
as
follows:
\begin{equation}
\begin{aligned}
\xi_A(r) &= \begin{pmatrix} a_i(r) \cr \alpha_\mu(r) \end{pmatrix} \,, & \qquad \xi^A(r) &= \xi_A(r)^\dag = \begin{pmatrix} a^i(r) \cr \alpha^\mu(r) \end{pmatrix} = \syng(1) \,,\\
\eta_A(r) &= \begin{pmatrix} b_i(r) \cr \beta_\mu(r) \end{pmatrix} \,, & \eta^A(r) &= \eta_A(r)^\dag = \begin{pmatrix} b^i(r) \cr \beta^\mu(r) \end{pmatrix} = \syng(1) \,,\\
\zeta_A & = \begin{pmatrix} c_i \cr \gamma_\mu \end{pmatrix} \,, & \zeta^A &= {\zeta_A}^\dag = \begin{pmatrix} c^i \cr \gamma^\mu \end{pmatrix} = \syng(1) \,,
\end{aligned}
\label{superoscillators}
\end{equation}
where the super-index
$A$ takes the values $1,2|1,2,3$ and $r=1,\dots,p$.
The nonvanishing super-commutation relations are
\begin{equation}
\left[ \xi_A(r) , \xi^B(s) \right\}   = \delta_A^B \delta_{rs} \,,
\qquad
\left[ \eta_A(r) , \eta^B(s) \right\} = \delta_A^B \delta_{rs} \,,
\qquad
\left[ \zeta_A , \zeta^B \right\}     = \delta_A^B \,,
\end{equation}
where the super-commutators are defined as
\begin{equation}
\left[ \xi_A(r) , \xi^B(s) \right\} = \xi_A(r) \xi^B(s) - (-1)^{(\deg A)(\deg B)} \xi^B(s) \xi_A(r) \,,
\end{equation}
etc., with $\deg A = 0$ ($\deg A = 1$) if $A$ is a bosonic (fermionic) index.

The $OSp(6|4)$ generators can then be realized as bilinears of these
superoscillators:
\begin{equation}
\begin{aligned}
S^{AB}    &= \vec{\xi}^A \cdot \vec{\eta}^B + \vec{\eta}^A \cdot \vec{\xi}^B + \epsilon \; \zeta^A \zeta^B
&&= \syng(2) \,,\\
S_{AB}    &= \vec{\xi}_A \cdot \vec{\eta}_B + \vec{\eta}_A \cdot \vec{\xi}_B + \epsilon \; \zeta_A \zeta_B
&&= (S^{BA})^\dagger \,,\\
M{}^A{}_B &= \vec{\xi}^A \cdot \vec{\xi}_B + (-1)^{(\deg A)(\deg B)} \vec{\eta}_B \cdot \vec{\eta}^A \\
&\quad + \frac{\epsilon}{2} \left( \zeta^A \zeta_B + (-1)^{(\deg A)(\deg B)} \zeta_B \zeta^A \right)
&&= (M{}^B{}_A)^\dagger \,.
\end{aligned}
\label{OSp64Rgenerators}
\end{equation}
Of course by restricting to purely bosonic or fermionic indices
we recover the generators
of the bosonic subgroups
\begin{equation}
\begin{aligned}
& (M{}^i{}_j = I{}^i{}_j, S_{ij} = K_{ij}, S^{ij} = P^{ij}) & \leftrightarrow &&& Sp(4,\mathbb{R}) \,,\\
& (M{}^\mu{}_\nu = U{}^\mu{}_\nu, S_{\mu\nu} = A_{\mu\nu}, S^{\mu\nu} = A^{\mu\nu}) & \leftrightarrow &&& SO(6) \,.
\end{aligned}
\end{equation}
Whenever the indices take specific integer values we will use the
notation of the previous sections for these bosonic generators.

The odd generators have one bosonic and one fermionic index,
and we will use the relation $(M{}^A{}_B)^\dagger = M{}^B{}_A$ to
always display the bosonic index to the left and
the fermionic index to the right.
The anticommutators among the odd generators are explicitly given by
\begin{equation}
\label{Osp64_anticommutators}
\begin{aligned}
\{ S_{i\mu} , S^{j\nu} \}       & =\delta_\mu^\nu I{}^j{}_i - \delta_i^j U{}^\nu{}_\mu \,,
&\qquad \{ S_{i\mu} , M{}^j{}_\nu \} &= -\delta_i^j A_{\mu\nu} \,,\\
\{ M{}^i{}_\mu , M{}_j{}^\nu \} & =\delta_\mu^\nu I{}^i{}_j + \delta_j^i U{}^\nu{}_\mu \,,
&\{ S_{i\mu} , M_j{}^\nu \} &= \delta_\mu^\nu K_{ij} \,,
\end{aligned}
\end{equation}
together with others obtained by hermitian conjugation.
Finally, the commutators between even and odd oscillators are
\begin{equation}
\begin{aligned}
\left[ I{}^i{}_j , M{}^k{}_\mu \right]
&= \delta_j^k M{}^i{}_\mu \,,
& \left[ U{}^\mu{}_\nu , M{}^k{}_\lambda \right]
&= -\delta^\mu_\lambda M{}^k{}_\nu \,,\\
\left[ I{}^i{}_j , M{}_k{}^\mu \right]
&= -\delta^i_k M{}_j{}^\mu\,,
& \left[ U{}^\mu{}_\nu , M{}_k{}^\lambda \right]
&= \delta^\lambda_\nu M{}_k{}^\mu \,,\\
\left[ I{}^i{}_j , S_{k\mu} \right]
&= -\delta^i_k S_{j\mu} \,,
& \left[ U{}^\mu{}_\nu , S_{k\lambda} \right]
&= -\delta_\lambda^\mu S_{k\nu} \,,\\
\left[ I{}^i{}_j , S^{k\mu} \right]
&= \delta_j^k S^{i\mu} \,,
& \left[ U{}^\mu{}_\nu , S^{k\lambda} \right]
&= \delta^\lambda_\nu S^{k\mu} \,,\\
\left[ K_{ij} , M{}^k{}_\mu \right]
&= \delta_i^k S_{j\mu} + \delta_j^k S_{i\mu} \,,
&\qquad \left[ A_{\mu\nu} , M{}^\lambda{}_k \right]
&= -\delta^\lambda_\mu S_{\nu k} + \delta^\lambda_\nu S_{\mu k} \,,\\
\left[ K_{ij} , S^{k\mu} \right]
&= \delta_i^k M{}_j{}^\mu + \delta_j^k M{}_i{}^\mu \,,
& \left[ A_{\mu\nu} , S^{k\lambda} \right]
&= -\delta^\lambda_\mu M{}^k{}_\nu + \delta^\lambda_\nu M{}^k{}_\mu \,,
\end{aligned}
\label{OSp64algebra}
\end{equation}
where we have again omitted commutators which can be obtained from these
by hermitian conjugation. From the above relations we can easily
determine the dilatation charges of the odd generators,
\begin{equation}
\left[ \Delta , J \right] = C(J)\,J \,,
\end{equation}
where $C(J)=\frac{1}{2}$ for $J=M^k{}_\mu, S^{k\mu}$ and
$C(J)=-\frac{1}{2}$ for $M{}_k{}^\mu, S_{k\mu}$, indicating that
these
correspond respectively
to the $6$ supersymmetry and $6$ superconformal
generators.

The procedure for the construction of $OSp(6|4)$ representations is a
refinement of what we saw for the bosonic sectors. We
define the vacuum $|0\rangle$ of the Fock space of states to
be the state which is annihilated by the covariant oscillators
$\xi_A(r),\eta_A(r), \zeta_A$, and then consider lowest-weight
states which
are annihilated by $S_{AB}$ and transform irreducibly under $U(2|3)$.
A non-exhaustive list of states that satisfy the first condition is
given by~\cite{Gunaydin:1985tc}
\begin{equation}
[\zeta^A]^{k_0} [\xi^B(1)]^{k_1} \cdots [\xi^C(r)]^{k_r} [\eta^D(r+1)]^{k_{r+1}} \cdots [\eta^E(p)]^{k_p} |0\rangle
\end{equation}
and
\begin{equation}
(\xi^A(r)\eta^B(r)-\eta^A(r)\xi^B(r)) |0\rangle \,,
\end{equation}
where $r=1,2,\ldots,p$, $k_0=0,1$ and the other
$k_i$ are nonnegative integers.
As we'll see in the next section, there exist a few more possibilities
for states that can be annihilated by $S_{AB}$, but these two types
capture the great majority of possible lowest-weight states.

The new elements when we look at supergroups come from the condition of
irreducibility, and from the fact that we are promoting the oscillators to
superoscillators. In particular, in order for linear combinations of
superoscillators to transform irreducibly under $U(2|3)$, it is necessary
to demand that if the superindices are symmetrized for bosonic values,
they have to be antisymmetrized for fermionic values, and vice versa.
This can be seen heuristically by convincing oneself that $\xi^A \xi^B$
transforms irreducibly, but when both indices are bosonic (fermionic)
it is automatically (anti-) symmetrized.

Thus we define graded symmetrization or
``supersymmetrization''
\begin{equation}
\xi^{(A} \eta^{B)} \equiv \xi^A \eta^B + \eta^A \xi^B
= \xi^A \eta^B + (-1)^{(\deg A) (\deg B)} \xi^B \eta^A \,,
\label{supersymmetrization}
\end{equation}
and graded antisymmetrization or ``superantisymmetrization''
\begin{equation}
\xi^{[A} \eta^{B]} \equiv \xi^A \eta^B -\eta^A \xi^B
= \xi^A \eta^B - (-1)^{(\deg A) (\deg B)} \xi^B \eta^A \,,
\label{superantisymmetrization}
\end{equation}
with the obvious extensions of these definitions to products of
more than two superoscillators. Products of
superoscillators with all their indices either supersymmetrized and/or
superantisymmetrized will form irreducible representations of
$U(2|3)$~\cite{Baha Balantekin:1980qy}, and these can categorized
in terms of super-Young tableaux (SYT) in the same manner that
symmetrizations and/or antisymmetrizations of $U(N)$ oscillators are
categorized by ordinary Young tableaux.

There exists a rich literature on super-Young
tableaux~\cite{Baha Balantekin:1980qy,Baha Balantekin:1980pp,Baha Balantekin:1981bk,Bars:1982se,Morel:1984vw,Hurni:1985vk}
and their
applications (see for
example~\cite{Gunaydin:1984wc,Gunaydin:1985tc,Gunaydin:1988kz,
Gunaydin:1998jc,Fernando:2004jt}),
so here we'll just
mention a few intuitive examples. To any contravariant superoscillator,
namely the ones of the second row of~(\ref{superoscillators}), there
corresponds single (slashed) SYT $\syng(1)$. For the two graded
symmetrized~(\ref{supersymmetrization}) and
antisymmetrized~(\ref{superantisymmetrization}) oscillators we have
the SYT
$\syng(2)$ and $\syng(1,1)$ respectively. Consequently $k$
oscillators of the same kind $\xi^{A_1}\cdots \xi^{A_k}$ will be
described by a SYT with
a single row of $k$ boxes, $\syng(1)\cdots\syng(1)$, and for
mixed products we first supersymmetrize all superoscillators corresponding
to each row, and then superantisymmetrize with respect to the columns.
All of these properties are analogous to ordinary Young tableaux, however a
major difference is unlike
ordinary Young tableaux a SYT can have any number or rows as a
consequence of the fact that graded antisymmetrization corresponds to
symmetrization of fermionic indices, and hence can be carried out
indefinitely.

Returning to our discussion of $OSp(6|4)$ representations, the lesson
is that we can obtain all states of a multiplet by tensoring the
SYT of its LWS with arbitrary supersymmetrized powers of
$S^{AB}=\syng(2)$. The SYT for the latter just follows by
comparing its oscillator form in~(\ref{OSp64Rgenerators})
with~(\ref{supersymmetrization}).
Since each $U(2|3)$ representation decomposes into a set of ``component''
representations of the bosonic subgroup $U(2)\times U(3)$, one way to
proceed with the tensoring is to first perform the decomposition and
then tensor the ordinary Young tableaux according to the usual rules.

We defer the details of the decomposition of $U(2|3)$ states to the
appendix and we provide information
about the tensoring procedure in section~\ref{partfn_method}.
Furthermore, in section~\ref{section_notations} we will give the
explicit relation between the SYT labeling of supermultiplets and
other widely used conventions.

\subsection{Serre-Chevalley Basis}
\label{section_chevalley_basis}

\FIGURE{
\epsfig{file=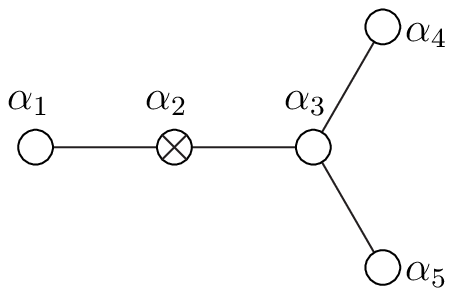}
\caption{The $OSp(6|4)$ Dynkin diagram.}
\label{DynkinDiagram}
}

In this section we study the structure of the $OSp(6|4)$ algebra and
determine the Cartan charges of each state in the oscillator construction.
As a useful application we also provide the representation labels for each
class of solutions of the ABJM theory's two-loop Bethe
ansatz~\cite{Minahan:2008hf}, a result also presented in~\cite{Minahan:2009te}
in a slightly different manner.

The $OSp(6|4)$ Dynkin diagram in the distinguished basis is
shown in figure~\ref{DynkinDiagram},
from which the Cartan matrix\footnote{We use the standard
convention where the nonzero diagonal elements are always equal to 2.
In many cases where the Cartan matrix appears in relation to the Bethe
ansatz, an alternative convention is also used where the rows
corresponding to the fermionic and conformal Cartan generators have their
signs flipped, see for example~\cite{Beisert:2003yb,Minahan:2008hf}.
This is permissible
due to the invariance of the Bethe ansatz under this inversion.}
\begin{equation}
{\cal{K}} = \left(\begin{array}{c|c|ccc}2& -1 &&&\\
\hline   -1 & &+1&&\\
\hline & -1 &+2&-1&-1\\
&&-1&+2&\\
&&-1&&+2
\end{array}\right)
\label{Cartan_matrix}
\end{equation}
follows. It is evident both from the Dynkin diagram and the Cartan matrix
that the roots $\alpha_3, \alpha_4, \alpha_5$ belong to the
$SO(6)\subset OSp(6|4)$ subgroup while $\alpha_1$ corresponds to the
$SU(2)\subset Sp(4,\mathbb{R}) \subset OSp(6|4)$ subgroup.

We would like to express the $OSp(6|4)$ superalgebra in a Serre-Chevalley
basis (see for example~\cite{Frappat:1996pb}), which is
defined by the relations
\begin{equation}
\begin{aligned}
\left[ H_i , H_j \right] &= 0 \,,\\
\left[ H_i , E^\pm_j \right] &= \pm {\cal{K}}_{ij} E^\pm_j \,,\\
\left[ E^+_i , E^-_j \right\} &= H_i \delta_{ij} \,,\\
\{ E^\pm_i , E^\pm_j \} &=0 & \text{if ${\cal{K}}_{ii}=0$} \,,\\
(ad_{E^\pm_i})^{1- {\cal{\tilde K}}_{ij}} E^\pm_j &=0 \,,
\end{aligned}
\label{Chevalleybasis}
\end{equation}
where
${\cal{K}}_{ij}$ are the matrix elements of the Cartan matrix
${\cal{K}}$, ${\cal{\tilde K}}=({\cal{\tilde K}}_{ij})$ is deduced
from ${\cal{K}}$ by replacing all its positive off-diagonal entries by
$-1$, and the last line means $(1-{\cal{\tilde K}}_{ij})$ times the
adjoint action of $E^\pm_i$ on $E^\pm_j$, which in turn is defined as
\begin{equation}
(ad_x) y   = \left[x,y\right\} \,, \qquad
(ad_x)^2 y = \left[x,\left[x,y\right\}\right\} \,, \qquad
\text{etc.}
\end{equation}

The type of the Dynkin diagram also requires supplementary
conditions to be imposed around the fermionic root
$\alpha_2$~\cite{Frappat:1996pb},
\begin{equation}
(ad_{E^\pm_2})(ad_{E^\pm_3})(ad_{E^\pm_2}) E^\pm_1
= (ad_{E^\pm_2})(ad_{E^\pm_1})(ad_{E^\pm_2}) E^\pm_3
= 0 \,.
\label{supplementarycond}
\end{equation}
Starting from~(\ref{Sp4Ralgebra}), (\ref{SO6algebra}), (\ref{Osp64_anticommutators})
and~(\ref{OSp64algebra}), the second and third relations
of~(\ref{Chevalleybasis}) together with the Cartan matrix
information~(\ref{Cartan_matrix}) are essentially sufficient for
determining $H_i$ and $E_i^\pm$, and then the remaining relations
of~(\ref{Chevalleybasis}) and~(\ref{supplementarycond}) can be readily
verified. In this way we find that the Cartan charges are given by
\begin{equation}
\begin{aligned}
H_1 &= I{}^2{}_2-I{}^1{}_1 = N_{B_2}-N_{B_1} \,,\\
H_2 &= I{}^1{}_1+U{}^3{}_3 = N_{B_1}+N_{F_3} \,,\\
H_3 &= U{}^3{}_3-U{}^2{}_2 = N_{F_3}-N_{F_2},\\
H_4 &= U{}^2{}_2-U{}^1{}_1 = N_{F_2}-N_{F_1} \,,\\
H_5 &= U{}^2{}_2+U{}^1{}_1 = N_{F_1}+N_{F_2}-f \,,
\end{aligned}
\end{equation}
whereas the corresponding raising/lowering operators are
\begin{equation}
\begin{aligned}
E_1^+ &= I{}^2{}_1 \,, & E_1^- &= I{}^1{}_2 \,,\\
E_2^+ &= M{}^1{}_3 \,, & E_2^- &= M{}_1{}^3 \,,\\
E_3^+ &= U{}^3{}_2 \,, & E_3^- &= U{}^2{}_3 \,,\\
E_4^+ &= U{}^2{}_1 \,, & E_5^- &= U{}^1{}_2 \,,\\
E_5^+ &= A^{21} \,, &\qquad E_4^- &= A_{12} \,,
\end{aligned}
\end{equation}
and we see that by construction $(E^+_i)^\dagger=E^-_i$. As an
independent check on this result
we considered the Chevalley bases of $Sp(4,\mathbb{R)}$
and $SU(4)$ separately and obtained the fermionic Cartan charge from them
according to~\cite{Hurni:1981yy}, finding agreement with the above
straightforward calculation.

Once we have the Serre-Chevalley basis it is easy to establish the
relation between the weight of any given state and its
excitation numbers (i.e., the numbers of
each type of raising operators needed to
reach it from a LWS).  In particular,
given that the LWS annihilated by all $E^-_i$ is
\begin{equation}
|\Omega\rangle \equiv \left( |0\rangle\tensor \gamma^1|0\rangle \right)^L \,,
\end{equation}
then an arbitrary state
\begin{equation}
(E^+_1)^{K_w}(E^+_2)^{K_s}(E^+_3)^{K_r}(E^+_4)^{K_v}(E^+_5)^{K_u}|\Omega\rangle
\end{equation}
(the choice of notation for the $K$'s
here anticipates the connection with that of~\cite{Minahan:2008hf})
will have number operator eigenvalues
\begin{equation}
\begin{aligned}
N_{B_1} &= K_s-K_w \,, & H_1 &= 2K_w-K_s \,,\\
N_{B_2} &= K_w \,, & H_2 &= K_r-K_w \,,\\
N_{F_1} &= L+K_u-K_v \,, & H_3 &= 2K_r-K_u-K_v-K_s \,,\\
N_{F_2} &= K_u+K_v-K_r \,, &\qquad H_4 &= 2K_v-K_r-L \,,\\
N_{F_3} &= K_r-K_s \,, & H_5 &= 2K_u-K_r-L \,.
\end{aligned}
\end{equation}

The right-hand sides of the above relations give us the weights of a
state with excitation numbers $\{K_w,K_s,K_r,K_v,K_u\}$. In particular
$H_1$ is twice the $SU(2)\subset Sp(4,\mathbb{R})$ spin and $[H_4,H_3,H_5]$
are the $SU(4)$ weights in the Dynkin basis.
Furthermore if we assume their values
refer to the the LWS of a certain $OSp(6|4)$ supermultiplet, then by
unitarity the labels of the corresponding HWS will
simply be\footnote{These expressions of the representation labels in terms
of number operators also make contact with and explain~(\ref{Sp4_labels})
and~(\ref{SU4_yt2dynk}). The value of $j$ is
determined by the state whose $U(2)$ YT has only oscillators with
index 1 on the first row and only oscillators with index 2 on the second
row. An analogous statement holds for the $SU(4)$ labels.}
$j=-H_1/2 = (N_{B_1} - N_{B_2})/2$
and $[d_1,d_2,d_3]=[-H_5,-H_3,-H_4] = [f-N_{F_1}-N_{F_2},
N_{F_2}-N_{F_3}, N_{F_1}-N_{F_2}]$. Together with
the scaling dimension~(\ref{Delta}) and the number of generations $f$,
these comprise the Dynkin labels of the supermultiplet in question, and
we have shown that they are related to the excitation numbers of the LWS as
\begin{equation}
\begin{aligned}
\Delta &= L+\frac{1}{2} K_s \,, & j &=\frac{1}{2} K_s-K_w \,, & f &= 2L \,,\\
d_1 &= L+K_r-2K_u \,, &\quad d_2 &= K_v+K_u-2K_r+K_s \,, &\quad d_3 & =L+K_r-2K_v \,,
\end{aligned}
\end{equation}
in exact agreement with~\cite{Minahan:2009te}. Since the excitation
numbers ${K}$ are the same quantities that appear in the theory's
Bethe ansatz~\cite{Minahan:2008hf} (denoting the number of Bethe roots
of each type which appear in a solution of the Bethe equations)
this formula is useful in identifying
which symmetry multiplet any particular solution belongs to.

\section{Representations and their Partition Functions}
\label{sectionthree}

\subsection{Notational Conventions}
\label{section_notations}

We have seen that each $OSp(6|4)$ supermultiplet can be characterized by the
number $f$ of generations of superoscillators used in realizing the
generators, see~(\ref{OSp64Rgenerators}), and the SYT
of its LWS.
We will use $\cv^f$ to denote the representation with $f$ generations
whose LWS is $|0\rangle$.
For a more general $OSp(6|4)$
supermultiplet whose LWS transforms in a non-trivial representation
of $U(2|3)$ we indicate the SYT of that $U(2|3)$ representation
as a subscript on $\cv^f$.
Thus the representation with $f=2$ and LWS
$\xi^A\cdots\xi^B|0\rangle=\syng(2)\cdots\syng(1)$
will be denoted ${\cal V}^2_{\Yboxdim5.7pt\styng(2)\cdots\styng(1)}$,
and so on.

The information provided graphically by the SYT can equally well be encoded
in labels $(k_1,k_2,\ldots)$ which indicate
the number of boxes
in the (first, second, $\ldots$) row of the tableau.
Note that since the number of rows is not fixed, as we saw in
section~\ref{sytsection},
the number of labels will also vary for a given $f$.
For example we can have $\V{f}{2}=\cv^f_2$, $\V{f}{3,2}=\cv^f_{3,2}$, etc.

More conventionally, a multiplet may alternatively be
described by the eigenvalues
of some set of states within the multiplet under the action of
the generators of the Cartan subalgebra.
The simplest choice is to use the labels of the bosonic subgroup
$Sp(4,\mathbb{R}) \times SO(6)$ we saw earlier, namely the scaling dimension
$\Delta$, the spin in a particular axis in
$SO(3)\subset SO(3,2)$, denoted by
$j$, and the $SU(4)$ Dynkin labels $[d_1,d_2,d_3]$.
As far as choosing a particular state whose charges will be used to label
the entire multiplet, we pick a primary state,
defined as a state annihilated by the superconformal generators,
or alternatively as a state with the lowest scaling dimension in the multiplet.
In order to have positive labels we also demand that the state is of
highest-weight type with respect to both $SO(3)\subset SO(3,2)$ and $SU(4)$.
However as we saw in section~\ref{section_chevalley_basis}
all the information can be obtained by the appropriate  LWS by unitarity.

So the translation between the SYT labeling and
the Cartan labeling of $OSp(6|4)$
supermultiplets simply consists of finding the $Sp(4,\mathbb{R})\times SO(6)$
submultiplet with the smallest
number of $U(2)\subset Sp(4,\mathbb{R})$ boxes in the respective decomposition
(see the appendix) and calculating its Cartan
labels with the help of~(\ref{Sp4_labels}) and~(\ref{SU4_yt2dynk}).
We mention them here combined for convenience:
\begin{equation}
\label{OSp64labels}
[\Delta,j;d_1,d_2,d_3]=
[\frac{1}{2} (m_1+m_2+f),\frac{1}{2}(m_1-m_2);f-l_1-l_2,l_2-l_3,l_1-l_2]\,.
\end{equation}
We will use $\Vd{\Delta, j}{d_1,d_2,d_3}{f}$
to denote the representation with the given labels. We will also use
a bar to denote the `conjugate' of a representation, which is
obtained by exchanging two of the $SU(4)$ labels:
$\Vdb{\Delta,j}{d_1,d_2,d_3}{f}=\Vd{\Delta, j}{d_3,d_2,d_1}{f}$.

The procedure of finding the $Sp(4,\mathbb{R})\times SO(6)$ submultiplet
with the lowest scaling dimension that we described above can in fact
be performed for an arbitrary SYT, thus yielding a general formula to
relate the Cartan labels of supermultiplet and its SYT labels
$(k_1,k_2,\ldots,k_n)$.  The result is
that, for $k_n\le\ldots\le k_3\le 3$,
\begin{equation}
V^f_{k_1,k_2,\ldots,k_n}=\Vd{\Delta, j}{d_1,d_2,d_3}{f}
\end{equation}
with
\begin{equation}
\begin{aligned}
\Delta&=\frac{1}{2}\left(\max(k_1-3,0)+\max(k_2-3,0)+f \right)\,,\\
j&=\frac{1}{2}(\max(k_1-3,0)-\max(k_2-3,0))\,,\\
d_1&=f-\sum_{i=1}^n \min(k_i,2)\,,\\
d_2&=\sum_{i=1}^n \delta_{k_i,2}\,,\\
d_3&=\sum_{i=1}^n \delta_{k_i,1}\,.
\end{aligned}
\label{yt2dynkin}
\end{equation}
For more information about the relation between SYT and Dynkin labels
see~\cite{Bars:1982se}.

Finally we will also make use of characters for $OSp(6|4)$ representations.
We define the characters to weigh states according to the
Cartan charges~(\ref{OSp64labels}), specifically
\begin{equation}
\chi_\cv(x,y,u,r,v)
=\text{Tr}_\cv\lbrack x^\Delta y^{j}u^{d_1}r^{d_2}v^{d_3}\rbrack\,,
\label{chidef}
\end{equation}
given by the trace
over all states of a particular $OSp(6|4)$ representation $\cv$.
Explicit formulas for all $OSp(6|4)$ characters may be found
in~\cite{Dolan:2008vc}, so in order to save space we will for the
most part only
display explicit formulas for the `partition functions'
\begin{equation}
\label{partfn_def}
V(x)=\text{Tr}_\cv\lbrack x^\Delta\rbrack=\chi_\cv(x,1,1,1,1)
\end{equation}
which count the number of states within the multiplet $\cv$ for each value
of the classical scaling dimension $\Delta$.

\subsection{Calculating Characters: An Example}
\label{partfn_method}

In this section we demonstrate by a
a particular example the steps for calculating
the character of a general $OSp(6|4)$ representation using the oscillator
construction. At the end of the day we will focus on the partition functions
defined in~(\ref{partfn_def}), and show how the general formulas reduce to
those. The reader interested in the final answer may jump to the next section,
where the most relevant partition functions are summarized.

The basic computational strategy exploits the fact that an $OSp(6|4)$
supermultiplet decomposes into multiplets of the bosonic subgroup
in order to express its character in terms of a sum of
$Sp(4,\mathbb{R})\times SO(6)$ characters. Due to the noncompactness of
$Sp(4,\mathbb{R})$ we also reduce the calculation of its character to a sum
of the characters of its maximal compact subgroup $U(2)$.
For simplicity we initially sum over the $U(2)$ multiplets that don't contain
spacetime derivatives (generated by $P^{ij}$), and account
for the missing
states only at the very end.

Finding the bosonic lowest-weight states
is a two-step process. First, we decompose the
LWS of the supermultiplet according to the rules tabulated
in the appendix. The occurring states are
annihilated by all of the
$S_{AB}$, so they certainly are lowest-weight
states of the $Sp(4,\mathbb{R})\times SU(4)$ bosonic subgroup
as well, but there are additional $Sp(4,\mathbb{R}) \times SU(4)$
lowest weight states which are annihilated by all of the
$K_{ij}$ and $A_{\mu\nu}$ but not by all of the $S_{AB}$.

We find the remaining $Sp(4,\mathbb{R})\times SU(4)$
lowest-weight states by acting on the supermultiplet LWS with the odd
generators $S^{i\mu}$. Once we have obtained all the lowest-weight states
of the bosonic subgroup with this method we can then find the labels
$[\Delta, s; d_1, d_2,d_3, f]$ of their corresponding submultiplets
with the help of~(\ref{yt2dynkin}).

As far as the action of the odd raising operators on the supermultiplet LWS
is concerned, we use the fact that the $S^{i\mu}$ transform in the
(fundamental,fundamental) of the $U(2) \times U(3)$ subgroup
of $Sp(4,\mathbb{R})\times SO(6)$, with corresponding YT
$\left(\yng(1),\yng(1)\right)$. Also, since the $S^{i\mu}$ generators
anticommute with each other we can have nonvanishing products of no more
than six of them, and those products will transform in
antisymmetrized powers of $\left(\yng(1),\yng(1)\right)$. We can thus write
\begin{equation}
\begin{aligned}
S^{i\mu}&=\left( \yng(1), \yng(1)\right)\,,\\
\left(S^{i\mu}\right)_-^{ 2}&=\left( \yng(2), \yng(1,1)\right)+\left( \yng(1,1), \yng(2)\right)\,,\\
\left(S^{i\mu}\right)_-^{ 3}&=\left( \yng(3), \yng(1,1,1)\right)+\left( \yng(2,1), \yng(2,1)\right)\,,\\
\left(S^{i\mu}\right)_-^{ 4}&=\left( \yng(3,1), \yng(2,1,1)\right)+\left( \yng(2,2), \yng(2,2)\right)\,,\\
\left(S^{i\mu}\right)_-^{5}&=\left( \yng(3,2), \yng(2,2,1)\right)\,,\\
\left(S^{i\mu}\right)_-^{ 6}&=\left( \yng(3,3), \yng(2,2,2)\right)\,,
\end{aligned}
\label{oddraising_ops}
\end{equation}
and the $Sp(4,\mathbb{R})\times SO(6)$ multiplets arising from the
action of the $S^{i\mu}$ generators on the $OSp(6|4)$ supermultiplet may
be deduced with the help of~(\ref{oddraising_ops}), the decompositions
of the appendix and the
usual rules for tensor products of $U(2)\times U(3)$ Young tableaux.

Finally, a subtlety that sometimes arises is that the $U(3)$ Young
tableaux resulting from the tensor products may lead to $SU(4)$ Dynkin
labels where one of the entries is negative, due to the formula
$d_1=f-l_1-l_2$. These representations may either vanish, or be mapped
to representations with positive labels, after reflecting the roots of
the $SU(4)$ group. The action of root reflections is encoded in
the corresponding Weyl group, which for $SU(n)$ groups is $S_n$,
the symmetric group on $n$ elements. In the case of $SU(4)$ we have $S_4$,
which is of order 24 and is generated by three
elementary reflections which act on the Dynkin labels as
\begin{equation}
\begin{aligned}
\sigma_1([d_1,d_2,d_3])&=[-d_1, d_1 + d_2, d_3]\,,\\
\sigma_2([d_1,d_2,d_3])&=[d_1+d_2, -d_2, d_2+d_3]\,,\\
\sigma_3([d_1,d_2,d_3])&=[d_1, d_2 + d_3, -d_3]\,,\\
\end{aligned}
\end{equation}
with the remaining elements given by various
products of these, e.g.~$\sigma_1 \sigma_3$, $\sigma_1 \sigma_2\sigma_1$, etc.
What is relevant in our case is the so-called ``shifted action'' of the
elements of the Weyl group, defined as
\begin{equation}
[d_1,d_2,d_3]^{\sigma_i}={\sigma_i}([d_1+1,d_2+1,d_3+1])-[1,1,1]\,,
\end{equation}
yielding
\begin{equation}
\begin{split}\label{shifted_action}
[d_1,d_2,d_3]^{\sigma_1}&=[-d_1-2,d_1+d_2+1,d_3]\\
[d_1,d_2,d_3]^{\sigma_2}&=[d_1+d_2+1,-d_2-2,d_2+d_3+1]\\
[d_1,d_2,d_3]^{\sigma_3}&=[d_1,d_2+d_3+1,-d_3-2],
\end{split}
\end{equation}

In order to determine whether a negative-labeled representation contributes
or not, we act on them with the ``shifted action''~(\ref{shifted_action}) of
all of the elements of the Weyl group. At most one of the elements, which
can be expressed as a product of $k$ elementary reflections, may render all
the Dynkin labels positive. If such an element exists
then this representation contributes to the tensor product with
the transformed labels and
an overall
sign $(-1)^k$.

If no Weyl group element can turn all the labels positive then the
representation does not contribute\footnote{This ``filtering'' of
negative-label representations to vanishing and contributing ones with the
combined action $(-1)^k [d_1,d_2,d_3]^{\sigma_i}$ is what is more generally
defined as the alternating Weyl sum, and it plays
an essential part of the Racah-Speiser algorithm for decomposing tensor
products into irreducible representations.  See~\cite{Fuchs:1997b} for a
general discussion, and the appendix of~\cite{Dolan:2002zh} for the
particular application to $SU(4)$. Here we have to consider this step
for $d_1$ separately as the $U(3)$-invariant method
of tensoring $S^{i\mu}$ with the LWS
does not address it automatically.}.
One can explicitly check that a representation will vanish if any of the
following conditions on the initial labels holds:
\begin{equation}
\begin{aligned}
d_1&=-1\,,&d_2&=-1\,,&d_3&=-1\,,\\
d_1+d_2&=-2\,,&d_2+d_3&=-2\,,&d_1+d_2+d_3&=-3\,.
\end{aligned}
\end{equation}

We will now illustrate how this procedure works by an explicit
calculation of the character for $\cv^2_1$. The supermultiplet LWS is
\begin{equation}
\label{oddraising_action1}
\xi^A |0\rangle=\syng(1)=(\overset{1/2}{\yng(1)},\overset{\mathbf{10}}{1})+(\overset{0}{1},\overset{\mathbf{15}}{\yng(1)})\,,
\end{equation}
where we have also included the corresponding $SU(2)$ Cartan
charge $j$ on top of the $U(2)$ YT and the corresponding $SU(4)$ multiplet
on top of the $U(3)$ YT\footnote{For compactness we indicate just the
dimensionality
of the representation as a proxy for its Dynkin labels: $\mathbf{10}=[2,0,0],
\mathbf{\overline{10}}=[0,0,2], \mathbf{15}=[1,0,1], \mathbf{6}=[0,1,0],
\mathbf{1}=[0,0,0]$.}.
Acting with one odd raising operator yields
\begin{equation}
\begin{aligned}
S^{i\mu}\, \xi^A |0\rangle&=\left[( \yng(2), \yng(1,1))+( \yng(1,1), \yng(2))\right]\tensor \left[(\yng(1),1)+(1,\yng(1))\right]\\
&=(\overset{1}{\yng(2)},\overset{\mathbf{15}}{\yng(1)})+(\overset{0}{\yng(1,1)},\overset{\mathbf{15}}{\yng(1)})+(\overset{1/2}{\yng(1)},\overset{\mathbf{\overline{10}}}{\yng(2)})+(\overset{1/2}{\yng(1)},\overset{\mathbf{6}}{\yng(1,1)})\,.
\end{aligned}
\end{equation}
Continuing with a second odd raising operator we find
\begin{equation}
\begin{aligned}
(S^{i\mu})^{\ 2}\, \xi^A |0\rangle &=(\overset{3/2}{\yng(3)},\overset{\mathbf{6}}{\yng(1,1)})+(\overset{1/2}{\yng(2,1)},\overset{\mathbf{6}}{\yng(1,1)})
+(\overset{1/2}{\yng(2,1)},\overset{\mathbf{\overline{10}}}{\yng(2)})+(\overset{1}{\yng(2)},\overset{\mathbf{1}}{\yng(1,1,1)})\\
&\qquad+
\cancel{(\yng(2),\yng(2,1))}+\cancel{(\yng(1,1),\yng(2,1))}+\cancel{({\yng(1,1)},\yng(3))},
\end{aligned}
\end{equation}
where the three last terms vanish because they all have
$d_1=f-l_1-l_2=-1$. From now on we will suppress such vanishing
representations in all tensor products.

Moving on we find that the $Sp(4,\mathbb{R})\times SO(6)$ lowest-weight
states arising from the action of more odd raising operators are
\begin{equation}
\label{oddraising_action5}
\begin{aligned}
(S^{i\mu})^{3}\, \xi^A |0\rangle &=(\overset{2}{\yng(4)},\overset{\mathbf{1}}{\yng(1,1,1)})+(\overset{1}{\yng(3,1)},\overset{\mathbf{1}}{\yng(1,1,1)})+(\overset{1/2}{\yng(2,1)},\overset{-\mathbf{6}}{\yng(2,2)})+(\overset{1/2}{\yng(2,1)},\overset{-\mathbf{\overline{10}}}{\yng(3,1)})
\,,\\
(S^{i\mu})^{4}\, \xi^A |0\rangle &=(\overset{1}{\yng(3,1)},\overset{-\mathbf{1}}{\yng(2,2,1)})+(\overset{1/2}{\yng(3,2)},\overset{-\mathbf{6}}{\yng(2,2)})
+(\overset{0}{\yng(2,2)},\overset{-\mathbf{1}}{\yng(2,2,1)})+(\overset{0}{\yng(2,2)},\overset{-\mathbf{15}}{\yng(3,2)}) \,,\\
(S^{i\mu})^{5}\, \xi^A |0\rangle &=(\overset{1}{\yng(4,2)},\overset{-\mathbf{1}}{\yng(2,2,1)})+(\overset{0}{\yng(3,3)},\overset{-\mathbf{1}}{\yng(2,2,1)}) \,,\\
(S^{i\mu})^{6}\, \xi^A |0\rangle &=(\overset{0}{\yng(3,3)},\overset{\mathbf{1}}{\yng(3,2,2)})\,,
\end{aligned}
\end{equation}
where in the $SU(4)$ dimensionality we have also included the overall
sign $(-1)^k$ coming from the action of the Weyl group elements. In
more detail, we have for the particular negative-labeled $U(3)$ Young
tableaux
\begin{equation}
\begin{aligned}
\yng(2,2): [d_1,d_2,d_3]&=[-2,2,0]\Rightarrow[d_1,d_2,d_3]^{\sigma_1}=[0,1,0],\quad (-1)^k=-1 \,,\\
\yng(3,1): [d_1,d_2,d_3]&=[-2,1,2]\Rightarrow[d_1,d_2,d_3]^{\sigma_1}=[0,0,2],\quad (-1)^k=-1 \,,\\
\yng(2,2,1): [d_1,d_2,d_3]&=[-2,1,0]\Rightarrow[d_1,d_2,d_3]^{\sigma_1}=[0,0,0],\quad (-1)^k=-1 \,,\\
\yng(3,2): [d_1,d_2,d_3]&=[-3,2,1]\Rightarrow[d_1,d_2,d_3]^{\sigma_1}=[1,0,1],\quad (-1)^k=-1 \,,\\
\yng(3,2,2): [d_1,d_2,d_3]&=[-3,0,1]\Rightarrow[d_1,d_2,d_3]^{\sigma_3 \sigma_1}=[0,0,0],\quad (-1)^k=1 \,.
\end{aligned}
\label{oddraising_action6}
\end{equation}

Now summing all
contributions~(\ref{oddraising_action1})-(\ref{oddraising_action5})
we find that the trace~(\ref{chidef})
(still restricted to the states in $\cv^2_1$
that do not contain any derivatives $P^{ij}$) is
\begin{equation}
\label{chi_tilde}
\begin{split}
\tilde \chi^2_1&=x^{3/2} \chi_{(1/2)}(\chi_{[200]}+\chi_{[002]} )\\
&\quad+\big((x +x^2-x^3) \chi_{(0)}+x^2 \chi_{(1)}\big)\chi_{[101]}\\
&\quad+ \big((x^{3/2}-x^{7/2})\chi_{(1/2)}+x^{5/2} \chi_{(3/2)}\big)\chi_{[010]}\\
&\quad+ \big(-x^3 \chi_{(0)}+(x^2-x^4) \chi_{(1)}+x^3 \chi_{(2)}\big)\chi_{[000]}\,,
\end{split}
\end{equation}
where
\begin{equation}
\chi(y)_{(j)}=\frac{y^{-j}(1-y^{1+2j})}{1-y}
\end{equation}
is the character of the $SU(2)\subset Sp(4,\mathbb{R})$
representation with spin $j$ and
$\chi_{[d_1 d_2 d_3]}(u,r,v)$ is the character of the $SU(4)$ representation
with Dynkin labels $[d_1,d_2,d_3]$. The latter can be obtained by
a slight
modification of the $U(4)$ character~\cite{Baha Balantekin:1980qy},
\begin{equation}
\chi(u,r,v)_{[d_1 d_2 d_3]}=
\frac{\det(\epsilon^{n_l+4-l}_i)}{\det(\epsilon^{4-l}_i)}
\end{equation}
where $i$ and $l$ label the rows and columns of a $4 \times 4$ matrix,
the $U(4)$ YT labels $n_l$ are written in terms of the Dynkin labels
according to
\begin{equation}
n_l=\sum_{i=l}^{3} d_i \qquad\text{for }l=1,2,3\,,\qquad n_4=0
\end{equation}
for $SU(4)$, and the variables $\epsilon_i$ are
given by
\begin{equation}
\epsilon_1=u\,,
\qquad
\epsilon_2=\frac{r}{u}\,,
\qquad
\epsilon_3=\frac{v}{r}\,,
\qquad
\epsilon_4=\frac{1}{v}
\end{equation}
so as to count units of the Cartan charges in the Dynkin basis.
We notice that some powers of $x$ in the character~(\ref{chi_tilde})
have negative
coefficients as a result of including the Weyl-transformed negative-label
representations in~(\ref{oddraising_action6}).
As explained in~\cite{Dolan:2002zh} these negative contributions
are necessary to properly account for states which
vanish due to equations of motion or due to conservation equations.
A nice feature of including these states in the counting is that the full
character, when states involving the spacetime
derivatives $P^{ij}$ are included,
can be calculated naively with derivatives treated as if they acted freely.
Since the derivatives have $\Delta=1$ and belong to the spin $j=1$
representation, summing over the set of states that arise by acting on an
initial state with any number of derivatives introduces a
multiplicative factor of $1/((1-x y)(1-x)(1-x/y))$ into the character.

Finally then we arrive at the full character for $\cv^2_1$,
\begin{align}\label{fullcharacter}
\chi^2_1=\frac{\tilde \chi^2_1}{(1-x)(1-x y)(1-x/y)}\,,
\end{align}
with $\tilde \chi^2_1$ given in~(\ref{chi_tilde}).
In order to obtain the partition functions defined in~(\ref{partfn_def})
all we need to do is set $u,r,v$ and $y$ to $1$,
in which case the $SU(2)$ and $SU(4)$ characters reduce to the
dimension formulas of the respective representations,
\begin{equation}
\label{char_to_dim}
\begin{aligned}
\chi(1)_{(j)}&=2j+1\\
\chi(1,1,1)_{[d_1 d_2 d_3]}&=\frac{1}{12} (d_1 + 1) (d_2 + 1) (d_3 + 1)(d_1 + d_2 + 2) (d_2 + d_3 + 2)(d_1 + d_2 + d_3 + 3)\,.
\end{aligned}
\end{equation}

\subsection{All Irreducible Multiplets with up to 4 Sites}\label{section_reps}

For a fixed value of $f$ there exist only certain states that
can be annihilated by all of the $S_{AB}$ and hence serve as
lowest-weight states for
irreducible $OSp(6|4)$ representations. In this section we
tabulate all possible representations with $f\le4$ together with their
partition functions, obtained with the method described in
the previous section.
We have checked that all of the partition functions presented here are
consistent with the general formulas
presented in~\cite{Dolan:2008vc}.

At $f=1$ we have only the two
fundamental representations of $OSp(6|4)$, called the `singletons',
which are conjugate to each other:
\begin{equation}
\left.\begin{aligned}
{\cal V}^1&=\Vd{\frac{1}{2},0}{1,0,0}{1}\\
\V{1}{1}&=\cv^1_1=\Vd{\frac{1}{2},0}{0,0,1}{1}=\overline{{\cal V}}^1
\end{aligned}\quad\right\}
\quad V^1=\overline{V}^1=\frac{4\sqrt{x}}{(1-\sqrt{x})^2}\,.
\label{partfn_singleton}
\end{equation}
As is well known, $\cv^1$ includes the scalars of the ABJM theory and
their supersymmetric partners, transforming respectively in the ${\bf 4}$ and
${\bf \overline 4}$ of $SU(4)$. Note that obviously two
conjugate representations will always have
the same multiplicity of states at each energy level
$\Delta$, and hence will have equal partition functions.

\DOUBLETABLE{
\begin{tabular}{|cl|} \hline
& \\
$1$ & $\cv^2 = \Vd{1,0}{2,0,0}{2}$ \\ [12pt]
$\syng(1,1)$ & $\cv^2_{1,1} = \Vd{1,0}{0,0,2}{2} = \ocv^2$ \\ [12pt]
$\syng(1)$ & $\cv^2_1 = \Vd{1,0}{1,0,1}{2}$ \\ [12pt]
$\syng(2)$ & $\cv^2_2 = \Vd{1,0}{0,1,0}{2}$ \\ [12pt]
$\overbrace{\styng(2)\cdots\styng(1)}^{k\ge\,3}$ & $\cv^2_k = \Vd{\frac{k-1}{2},\frac{k-3}{2}}{0,0,0}{2}$ \\ [12pt] \hline
\end{tabular}
\label{f=2reps}
}{
\begin{tabular}{|cl|} \hline
& \\
$1$ & $\cv^3 = \Vd{\frac{3}{2},0}{3,0,0}{3}$ \\ [12pt]
$\syng(1,1,1)$ & $\cv^3_{1,1,1} = \Vd{\frac{3}{2},0}{0,0,3}{3} = \ocv^3$ \\ [12pt]
$\syng(1)$ & $\cv^3_1 = \Vd{\frac{3}{2},0}{2,0,1}{3}$ \\ [12pt]
$\syng(1,1)$ & $\cv^3_{1,1} = \Vd{\frac{3}{2},0}{1,0,2}{2} = \ocv^3_1$ \\ [12pt]
$\syng(2)$ & $\cv^3_2 = \Vd{\frac{3}{2},0}{1,1,0}{3}$ \\ [12pt]
$\syng(2,1)$ & $\cv^3_{2,1} = \Vd{\frac{3}{2},0}{0,1,1}{3} = \ocv^3_2$ \\ [12pt]
$\overbrace{\styng(2)\cdots\styng(1)}^{k\ge\,3}$ & $\cv^3_k = \Vd{\frac{k}{2},\frac{k-3}{2}}{1,0,0}{3}$ \\ [12pt]
$\overbrace{\styng(2,1)\raise 3.2pt\hbox{\,$\cdots$\styng(1)}}^{k\ge\,3}$ & $\cv^3_{k,1} = \Vd{\frac{k}{2},\frac{k-3}{2}}{0,0,1}{3} = \ocv^3_k$ \\ [12pt] \hline
\end{tabular}
\label{f=3reps}
}{The $f=2$ $OSp(6|4)$ multiplets.}{The $f=3$ $OSp(6|4)$ multiplets.}

Tables 1 and 2 display the possible multiplets for $f=2$ and $f=3$
respectively.
The partition functions of these representations are given by
\begin{equation}
\begin{aligned}
V^2 = V^2_{1,1} &= \frac{2x(5-x)}{(1-\sqrt{x})^3} \,,\\
V^2_1 &= \frac{x(15+7\sqrt{x}-3x-3x^{\frac{3}{2}})}{(1-\sqrt{x})^3} \,,\\
V^2_k &= x^{\frac{k-1}{2}}\frac{(1+\sqrt{x})^3}{(1-\sqrt{x})^3}\left(k-2+6\sqrt{x}-(k+2)x\right) \qquad k \ge 2 \,,
\end{aligned}
\end{equation}
and
\begin{equation}
\begin{aligned}
V^3 = V^3_{1,1,1} &= \frac{4x^{\frac{3}{2}}(5+3\sqrt{x})}{(1-\sqrt{x})^3} \,,\\
V^3_1 = V^3_{1,1} &= \frac{4x^{\frac{3}{2}}(9+11\sqrt{x}+4x)}{(1-\sqrt{x})^3} \,,\\
V^3_k = V^3_{k,1} &= \frac{4x^{\frac{k}{2}}(1+\sqrt{x})^3\left(k-2+(3+k)\sqrt{x}\right)}{(1-\sqrt{x})^3} \qquad k \ge 2 \,.
\label{vthreecharacters}
\end{aligned}
\end{equation}

\TABLE{
\begin{tabular}{|clccccl|} \hline
&&&&&& \\
$1$ & $\cv^4 = \Vd{2,0}{4,0,0}{}$ &&&& $\syng(1,1,1,1)$ & $\cv^4_{1,1,1,1} = \Vd{2,0}{0,0,4}{4} = \ocv^4$ \\ [12pt]
$\syng(1)$ & $\cv^4_1 = \Vd{2,0}{3,0,1}{4}$ &&&& $\syng(1,1,1)$ & $\cv^4_{1,1,1} = \Vd{2,0}{1,0,3}{2} = \ocv^4_1$ \\ [12pt]
$\syng(2)$ & $\cv^4_{2} = \Vd{2,0}{2,1,0}{4}$ &&&& $\syng(2,1,1)$ & $\cv^4_{2,1,1} = \Vd{2,0}{0,1,2}{4} = \ocv^4_2$ \\ [12pt]
$\overbrace{\styng(2)\cdots\styng(1)}^{k\ge\,3}$ & $\cv^4_k = \Vd{\frac{k+1}{2},\frac{k-3}{2}}{2,0,0}{4}$ &&&& $\overbrace{\styng(2,1,1)\raise 6.7pt\hbox{\,$\cdots$\styng(1)}}^{k\ge\,3}$ & $\cv^4_{k,1,1} = \Vd{\frac{k+1}{2},\frac{k-3}{2}}{0,0,2}{4} = \ocv^4_k$ \\ [12pt]
$\syng(1,1)$ & $\cv^4_{1,1} = \Vd{2,0}{2,0,2}{4}$ &&&& $\overbrace{\styng(2,1)\raise 3.2pt\hbox{\,$\cdots$\styng(1)}}^{k\ge\,3}$ & $\cv^4_{k,1} = \Vd{\frac{k+1}{2},\frac{k-3}{2}}{1,0,1}{4}$ \\ [12pt]
$\syng(2,1)$ & $\cv^4_{2,1} = \Vd{2,0}{1,1,1}{4}$ &&&& $\overbrace{\styng(2,2)\raise 3.2pt\hbox{\,$\cdots$\styng(1)}}^{k\ge\,3}$ & $\cv^4_{k,2} = \Vd{\frac{k+1}{2},\frac{k-3}{2}}{0,1,0}{4}$ \\ [12pt]
$\syng(2,2)$ & $\cv^4_{2,2} = \Vd{2,0}{0,2,0}{4}$ &&&& $\overbrace{\underbrace{\styng(1,1)\cdots\styng(1,1)}_{k_2\ge 3}\raise 3.2pt\hbox{\,$\cdots$\styng(1)}}^{k_1\ge k_2}$ & $\cv^4_{k_1,k_2} = \Vd{\frac{k_1+k_2-2}{2},\frac{k_1-k_2}{2}}{0,0,0}{4}$ \\ [12pt] \hline
\end{tabular}
\caption{The $f=4$ $OSp(6|4)$ multiplets.}
\label{f=4reps}
}
The analysis for $f=4$ is slightly more complicated since here
the number of boxes in the second row of the SYT can
be arbitrarily large.  The possible representations
are shown in table 3, while their partition functions are
\begin{equation}
\begin{aligned}
V^4 = V^4_{1,1,1,1} &= \frac{x^2(35+35\sqrt{x}+9x+x^\frac{3}{2})}{(1-\sqrt{x})^3} \,,\\
V^4_1 = V^4_{1,1,1} &= \frac{2x^2(35+59\sqrt{x}+36x+9x^\frac{3}{2}+x^2)}{(1-\sqrt{x})^3} \,,\\
V^4_{1,1} &= \frac{x^2(84+156\sqrt{x}+111x+39x^\frac{3}{2}+9x^2+x^\frac{5}{2})}{(1-\sqrt{x})^3}
\end{aligned}
\end{equation}
and
\begin{equation}
\begin{aligned}
V^4_k = V^4_{k,1,1} &= x^\frac{k+1}{2}\frac{(1+\sqrt{x})^3}{(1-\sqrt{x})^3}[10(k-2)+(k+1)(15\sqrt{x}+6x+x^\frac{3}{2})] \,,\\
V^4_{k,1} &= x^\frac{k+1}{2}\frac{(1+\sqrt{x})^3}{(1-\sqrt{x})^3}[15(k-2)+2(13k+6)\sqrt{x}+2(8k+1)x+6k x^\frac{3}{2}+k x^2] \,,\\
V^4_{k,2} &= x^\frac{k+1}{2}\frac{(1+\sqrt{x})^3}{(1-\sqrt{x})^3}[6(k-2)+4(4k-3)\sqrt{x}+(k-1)(20x+15x^\frac{3}{2}+6x^2+x^\frac{5}{2})] \,,\\
V^4_{k_1,k_2} &= (k_1-k_2+1)x^{\frac{k_1+k_2-2}{2}}\frac{(1+\sqrt{x})^9}{(1-\sqrt{x})^3} \,,
\end{aligned}
\end{equation}
where $k \ge 2$ and $k_1 \ge k_2 \ge 3$.

\section{Tensor Product Decompositions}
\label{tensorsection}

In this section we compute tensor product decompositions for products
of up to four copies of the $OSp(6|4)$ singletons $\cv^1$, $\ocv^1 = \cv^1_1$.
Before proceeding let us mention that although some of the results
here are not immediately obvious, the correctness of all of the decompositions
tabulated here can be verified with certainty using
$OSp(6|4)$ characters and the relation
$\chi_{\cv_a \tensor \cv_b}
= \chi_{\cv_a} \chi_{\cv_b}$.
We have performed this check using characters
constructed according to the procedure outlined in section~\ref{partfn_method}
or equivalently the expressions presented in~\cite{Dolan:2008vc}.

\subsection{Digraphs and Syllables of the ABJM Language}

Continuing the linguistic analogy mentioned in the introduction
we can think of the
irreducible representations arising in the tensor product of two
singletons as the `digraphs' (see also~\cite{Spradlin:2004pp})
of the ABJM language,
groups of two successive letters whose
phonetic value is a distinct sound, such as {\it aw} in {\it saw}
or {\it qu} in {\it question}. We have found
the consonant-vowel digraphs of the ABJM language to be
\begin{equation}
\cv^1\tensor\ocv^1=\sum_{m=0}^{\infty}\cv^2_{2m+1}=\Vd{1,0}{1,0,1}{2}+\sum_{n=0}^{\infty}\Vd{n+1,n}{0,0,0}{2} \,,
\label{f=2decomp1}
\end{equation}
and we also mention the decomposition of
a singleton squared, the consonant-consonant digraphs:
\begin{equation}
\cv^1\tensor\cv^1=\sum_{m=0}^{\infty}\cv^2_{2m}=\Vd{1,0}{2,0,0}{2}+\Vd{1,0}{0,1,0}{2}+\sum_{n=0}^{\infty}\Vd{n+\frac{3}{2},n+\frac{1}{2}}{0,0,0}{2} \,,
\label{f=2decomp2}
\end{equation}
with $\ocv^1\tensor\ocv^1$ simply being given by
the conjugate of the latter.

Taking the analogy with linguistics further we can refer
to the multiplets
appearing in triple singleton products, on which the Hamiltonian
density acts, as syllables, being the building blocks of words.
Of most interest are the consonant-vowel-consonant syllables given by
\begin{multline}
\cv^1\tensor\ocv^1\tensor\cv^1
= \sum_{m=0}^{\infty}(m+1)\left(\cv^3_{2m+1}+\cv^3_{2m+2,1}\right) \\
= \Vd{\frac{3}{2},0}{2,0,1}{3}+\Vd{\frac{3}{2},0}{0,1,1}{3}+\sum_{n=0}^{\infty}(n+2)(\Vd{n+\frac{3}{2},n}{1,0,0}{3}+\Vd{n+2,n+\frac{1}{2}}{0,0,1}{3})
\label{f=3decomp1}
\end{multline}
and we also mention the decomposition of a singleton cubed
\begin{equation}
\begin{aligned}
\cv^1\tensor\cv^1\tensor\cv^1
&= \sum_{m=0}^{\infty}(m+1)\left(\cv^3_{2m}+\cv^3_{2m+3,1}\right) \\
&= \Vd{\frac{3}{2},0}{3,0,0}{3}+2\Vd{\frac{3}{2},0}{1,1,0}{3} \\
&\quad +\sum_{n=0}^{\infty}\left[(n+3)\Vd{n+2,n+\frac{1}{2}}{1,0,0}{3}+(n+1)\Vd{n+\frac{3}{2},n}{0,0,1}{3}\right],
\end{aligned}
\label{f=3decomp2}
\end{equation}
again with the results for $\ocv^1 \tensor \cv^1 \tensor \ocv^1$ and
$\ocv^1\tensor\ocv^1\tensor\ocv^1$ obviously obtained by conjugation.

\subsection{Four-Fold Tensor Products}

Although the three-fold tensor product decompositions presented
above are sufficient for
analyzing the two-loop
Hamiltonian density,
physical states of the ABJM spin chain must have an even number of sites to be
gauge invariant so the shortest nontrivial `words' of the ABJM language
have length four.

The study of how these four-letter words arrange themselves into
irreducible multiplets of the theory's $OSp(6|4)$ therefore deserves
inquiry in its own right. This decomposition is more involved and
we proceed by splitting
the calculation into three steps. First we perform the
decomposition of only three out of the four sites,
\begin{equation}
\begin{aligned}
\cv^1\tensor\ocv^1\tensor\cv^1\tensor\ocv^1
&= (\cv^1\tensor\ocv^1\tensor\cv^1)\tensor\ocv^1 \\
&= \sum_{m=0}^{\infty}(m+1)\left(\cv^3_{2m+1}\tensor \ocv^1 +\cv^3_{2m+2,1}\tensor \ocv^1\right),
\end{aligned}
\label{4to3x1}
\end{equation}
and then we perform the decomposition between the $f=3$ irreducible
multiplets $\cv^3_{k}$ and $\cv^3_{k,1}$ with $\ocv^1$, for
which we find
\begin{equation}
\begin{aligned}
\cv^3_{2n+1}\tensor\ocv^1
&= \sum_{m=0}^\infty \sum_{j=0}^n \cv^4_{2n+2m+2,2j}+\cv^4_{2n+2m+1,2j+1} \,,\\
\cv^3_{2n}\tensor\ocv^1
&= \sum_{m=0}^\infty \left( \sum_{j=0}^n \cv^4_{2n+2m+1,2j}+ \sum_{j=0}^{n-1}\cv^4_{2n+2m,2j+1}\right),\\
\cv^3_{2n+1,1}\tensor\ocv^1
&= \sum_{m=0}^\infty\left( \ocv^4_{2n+2m+1}+ \sum_{j=1}^n \cv^4_{2n+2m+1,2j}+ \sum_{j=1}^{n+1}\cv^4_{2n+2m+2,2j-1}\right),\\
\cv^3_{2n,1}\tensor\ocv^1
&= \sum_{m=0}^\infty \left(\ocv^4_{2n+2m}+ \sum_{j=1}^n ( \cv^4_{2n+2m,2j}+ \cv^4_{2n+2m+1,2j-1} ) \right).
\end{aligned}
\label{3x1}
\end{equation}
One can also combine the results for odd and even multiplet indices if
desired,
\begin{equation}
\begin{aligned}
\cv^3_{k}\tensor\ocv^1
&= \sum_{m=0}^\infty \left( \sum_{j=0}^{[\frac{k}{2}]} \cv^4_{k+2m+1,2j}+ \sum_{j=0}^{[\frac{k-1}{2}]}\cv^4_{k+2m,2j+1}\right),\\
\cv^3_{k,1}\tensor\ocv^1
&= \sum_{m=0}^\infty \left(\ocv^4_{2n+k}+ \sum_{j=1}^{[\frac{k}{2}]} \cv^4_{2n+k,2j}+ \sum_{j=1}^{[\frac{k+1}{2}]}\cv^4_{2n+k+1,2j-1}\right).
\end{aligned}
\end{equation}
Finally we have to extract the overall coefficient of each
$V^4_{k_1,k_2}$ appearing in~(\ref{4to3x1}),
which is facilitated by the fact that the coefficients
of all the multiplets appearing in the direct sums~(\ref{3x1}) are all
equal to one. A closer inspection reveals that $\cv^4_{2j,2p}$ and
$\cv^4_{2j-1,2p-1}$ appear respectively in
\begin{itemize}
\item{$\cv^3_{2n+1}\tensor\ocv^1$ for $p\le n\le j-1$
and $p-1\le n\le j-1$, and}
\item{$\cv^3_{2n+2,1}\tensor\ocv^1$ for $p-1\le n\le j-1$
and $p-1\le n\le j-2$.}
\end{itemize}
It is clear from~(\ref{4to3x1})
that each $f=4$ multiplet in question will receive a
contribution of $(n+1)$ to its coefficient for each
particular
$\cv^3_{2n+1}\tensor\ocv^1$ or $\cv^3_{2n+2,1}\tensor\ocv^1$ in which
it is contained, yielding in total
the coefficients
\begin{equation}
\begin{aligned}
&\cv^4_{2j}, \ocv^4_{2j}:
&& \sum_{n=0}^{j-1}(n+1)=\frac{1}{2} j (j+1) \,,\\
&\cv^4_{2j,2p}:
&& \sum_{n=p}^{j-1}(n+1)+\sum_{n=p-1}^{j-1}(n+1)=j^2+j-p^2 \,,\\
&\cv^4_{2j-1,2p-1}:
&& \sum_{n=p-1}^{j-1}(n+1)+\sum_{n=p-1}^{j-2}(n+1)=j^2-p^2+p \,.
\end{aligned}
\end{equation}
So putting everything together we find the following decomposition
for the four-fold product of most interest,
\begin{multline}
\cv^1\tensor\ocv^1\tensor\cv^1\tensor\ocv^1
= \sum_{j=1}^\infty \frac{1}{2} j(j+1)\left(\cv^4_{2j}+\ocv^4_{2j}\right) \\
+ \sum_{j=1}^\infty \sum_{p=1}^j \left\{\left[j(j+1)-p^2\right] \cv^4_{2j,2p}+\left[j^2-p(p-1)\right]\cv^4_{2j-1,2p-1}\right\}.
\label{f=4decomp1}
\end{multline}
For completeness we also mention
the remaining four-fold product decompositions,
which can be calculated in a
similar fashion:
\begin{multline}
\cv^1\tensor\cv^1\tensor\cv^1\tensor\ocv^1
= \sum_{j=1}^\infty \frac{1}{2} j(j+1)\left(\cv^4_{2j-1}+\ocv^4_{2j+1}\right) \\
+ \sum_{j=1}^\infty \sum_{p=1}^j \left\{\left[(j+1)^2-p^2\right] \cv^4_{2j+1,2p}+\left[j(j+1)-p(p-1)\right]\cv^4_{2j,2p-1}\right\},
\label{f=4decomp2}
\end{multline}
and
\begin{multline}
\cv^1\tensor\cv^1\tensor\cv^1\tensor\cv^1
= \sum_{j=1}^\infty \frac{1}{2} j(j+1)\left(\cv^4_{2j-2}+\ocv^4_{2j+2}\right) \\
+ \sum_{j=1}^\infty \sum_{p=1}^{j} \left\{\left[j(j+1)+1-p^2\right] \cv^4_{2j,2p}+\left[j^2-1-p(p-1)\right]\cv^4_{2j-1,2p-1}\right\}.
\label{f=4decomp3}
\end{multline}
Notice that the very last term of the last relation actually has a
vanishing coefficient for $j=p=1$, but we have written it like this to
coincide with the previous term's summation range.
All other four-fold products may be obtained from~(\ref{f=4decomp1}),
(\ref{f=4decomp2}) or~(\ref{f=4decomp3}) by conjugation.

\subsection{Four-Letter Words of the ABJM Language}

In the spin chain description of gauge
theories only cyclically invariant spin chain states correspond
to gauge-invariant, single-trace operators.
For an ABJM spin chain with $f=2L$ sites
the physical states are those in the $+1$ eigenspace of the
projection operator
\begin{equation}
{\cal{P}}=\frac{1}{L}(1+T+T^2+\cdots+T^{L-1})=\frac{1}{L}\sum_{k=0}^{L-1}T^k \,,
\end{equation}
expressed in terms of the
translation operator $T$ which sends site $i$
to site $i+2$,
\begin{equation}
T|A_1 B_1 A_2\cdots A_L B_L\rangle=
(-1)^{\deg(A_1 B_1) \deg(A_2 B_2 \cdots A_L B_L)}|A_2 B_2 \cdots A_L B_L A_1 B_1\rangle \,,
\end{equation}
and obviously satisfies $T^L=1$.

If we focus on the
simplest nontrivial case $f=2L=4$, then ${\cal{P}}=\frac{1}{2}(1+T)$
and the physical subspace simply corresponds to the $+1$ eigenspace of $T$,
which in turn consists of states $|w\rangle$
which are symmetric in the permutation of
next-to-adjacent sites,
\begin{equation}
T(|A_1 B_1 A_2 B_2\rangle+s |A_2 B_2 A_1 B_1\rangle)
= +1(|A_1 B_1 A_2 B_2\rangle+s |A_2 B_2 A_1 B_1\rangle) \,,
\end{equation}
where $s = (-1)^{\deg(A_1 B_1) \deg(A_2 B_2)}$.
Hence the set of physical states will be given by the symmetric square of
$\cv^1\tensor\ocv^1$, which can further be expressed as
\begin{equation}
(\cv^1\tensor\ocv^1\tensor\cv^1\tensor\ocv^1)_+
= (\cv^1)^2_+ \tensor (\ocv^1)^2_++(\cv^1)^2_- \tensor (\ocv^1)^2_- \,,
\label{sym_first}
\end{equation}
in terms of symmetric and antisymmetric 2-fold tensor products
of $\cv$ and $\ocv$.

To proceed with the calculation of the symmetric and antisymmetric squares
we use the general character formula
(see for example~\cite{Frappat:2000a})
\begin{equation}
\chi_{(\cv)^2_\pm}(g)=\frac{1}{2} [(\chi_\cv (g))^2\pm \chi_\cv (g^2)] \,.
\label{character_formula}
\end{equation}
As pointed out in~\cite{Baha Balantekin:1980qy} this formula
still holds for supergroups if we use
supercharacters,
defined by including $(-1)^F$ inside the trace,
\begin{equation}
\chi^S_\cv(g)
= \text{Tr}_\cv((-1)^F g)\,.
\end{equation}

The oscillator construction makes clear a simple relation between
$\chi^S_\cv$ and $\chi_\cv$ for any $\cv$.  This hinges
on the observations
that $(-1)^F$ is (perhaps confusingly) just $(-1)^{N_B}$, where $N_B$
is the total boson number operator of~(\ref{boson_number_op}), and
that the bosonic and fermionic fields
have $\Delta$ equal to $\frac{1}{2}$ and 1 respectively.
Therefore
from~(\ref{Delta})
we see that
\begin{equation}
(-1)^F x^\Delta
= (-1)^{2\Delta - f} x^\Delta
= (-1)^f (-\sqrt{x})^{2\Delta}
\end{equation}
so that
\begin{equation}
\chi^S_\cv(x^\Delta)
= \chi^S_\cv((\sqrt{x})^{2\Delta})
= (-1)^f \chi_\cv ((-\sqrt{x})^{2\Delta}) \,.
\label{prescription_supercharacter}
\end{equation}
This relation gives a general prescription for obtaining
all supercharacters from the partition functions that we have already
calculated
in section~\ref{section_reps}\footnote{Note that
in~(\ref{prescription_supercharacter}) $x$ denotes whatever is
argument of the character that is exponentiated by $\Delta$. When
for example the argument is $x^2$, we would have to replace $x\to-x$.}.

Thus it is straightforward to
apply the character formula~(\ref{character_formula})
in order to calculate the symmetric and antisymmetric squares of $\cv^1$,
for which we find
\begin{equation}
(\cv^1)^2_+ = \sum_{m=0}^\infty \cv^2_{4m} \,,
\qquad
(\cv^1)^2_- = \sum_{m=0}^\infty \cv^2_{4m+2} \,,
\end{equation}
with the respective relations for $\ocv^1$ obtained from these by
conjugation (note
that $\ocv^2_{2j}=\cv^2_{2j}$ unless $j=0$).
Clearly the next step involves decomposing tensor products of the form
$\cv^2_{2l}\tensor \ocv^2_{2m}$, for which we find
\begin{equation}
\begin{aligned}
\cv^2_0\tensor\ocv^2_0
&= \sum_{p=0}^\infty \sum_{j=p}^\infty \cv^4_{2j+1,2p+1} \,,\\
\cv^2_0\tensor\ocv^2_{2m}
&= \sum_{p=0}^\infty \sum_{j=p+m}^{\infty} \cv^4_{2j+1,2p+1}+\cv^4_{2j,2p} \,,\\
\cv^2_{2l}\tensor\ocv^2_{2m}
&= \sum_{j=l+m}^\infty (\cv^4_{2j}+\ocv^4_{2j}) \,,\\
&\quad +\sum_{p=1}^m \sum_{j=l+m-p}^{\infty} c(j-l-m-p)\cv^4_{2j,2p}+c(j+1-l-m-p)\cv^4_{2j+1,2p-1}\\
&\quad +\sum_{p=m+1}^\infty \sum_{j=l+p-m}^{\infty} c(j-l-m-p)(\cv^4_{2j,2p}+\cv^4_{2j-1,2p-1}) \,,
\end{aligned}
\end{equation}
where $l\ge m\ge1$ and
\begin{equation}
c(k)=1+\Theta(k)=\begin{cases}
1& \text{if $k<0$} \,,\\
2& \text{if $k\ge0$} \,,
\end{cases}
\label{sym_last}
\end{equation}
and we use $\Theta(k)$ to denote the unit step function.

Combining all of these intermediate
steps we deduce that the set of
physical states for the $f=2L=4$ spin chain decomposes into irreducible
$OSp(6|4)$ multiplets as follows:
\begin{equation}
\begin{aligned}
(\cv^1\tensor\ocv^1)^2_+&=\sum_{j=1}^\infty j(j+1)(\cv^4_{4j}+\ocv^4_{4j}+\cv^4_{4j+2}+\ocv^4_{4j+2})+[2j(j-1)+1]\cv^4_{4j-3,1} \\
&\quad + \sum_{j=1}^\infty \sum_{p=1}^j
\left\{2\left[j(j+1)-p^2\right](\cv^4_{4j,4p}+\cv^4_{4j+2,4p})\right. \\
&\quad\quad\qquad\qquad +2\left[j^2-p(p-1)\right](\cv^4_{4j-1,4p-1}+\cv^4_{4j-1,4p-3}) \\
&\quad\quad\qquad\qquad+\left[2j^2-1-2p(p-1)\right] (\cv^4_{4j-2,4p-2}+\cv^4_{4j,4p-2}) \\
&\quad\quad\qquad\qquad\left.+ \left[2j(j+1)+1-2p^2\right] (\cv^4_{4j+1,4p-1}+\cv^4_{4j+1,4p+1}) \right\}.
\end{aligned}
\label{f=4sym_decomp}
\end{equation}

\section{The $OSp(4|2)$ Subsector}
\label{subsectorsection}

In this section we tabulate for completeness various results from the previous
two sections for the case of the $OSp(4|2) \subset OSp(6|4)$ subgroup.
This is of particular interest since the
two-loop dilatation operator has been constructed
explicitly in terms of its action on the fundamental
fields in this subsector
of the ABJM theory by Zweibel~\cite{Zwiebel:2009vb}.

The oscillator construction we reviewed
in section~\ref{section_oscillator_construction}
may be restricted to the $OSp(4|2)$ subsector simply by
restricting the bosonic and
fermionic oscillator indices to $i=1$ and $\mu=1,2$ respectively,
which gives the bosonic subgroup
$Sp(2,\mathbb{R})\times SO(4)$.  The
$Sp(2,\mathbb{R})\simeq SU(2)\simeq SO(3)$ content
of a LWS is characterized
just by its scaling dimension $\Delta=\frac{1}{2} (N_{B_1}+f)$, and
the $SO(4)\simeq SO(3)\times SO(3)$ content
is characterized by the Dynkin labels $[p,q]$
which are simply the Cartan charges of each of the two $SO(3)$s. In each
$Sp(2,\mathbb{R})$ multiplet there exists now only a single state for each
value of $\Delta$, whereas the dimensionality of an $SO(4)$
multiplet $[p,q]$ is $(p+1)(q+1)$. Similarly to (\ref{yt2dynkin}), the $OSp(4|2)$ multiplet with SYT labels $(k_1,...,k_n)$ will have Cartan labels
\begin{equation}
\begin{aligned}
\Delta&=\frac{1}{2}\left(\max(k_1-2,0)+f \right)\,,\\
p&=f-\sum_{i=1}^n \min(k_i,2)\,,\\
q&=\sum_{i=1}^n \delta_{k_i,1}\,.
\end{aligned}
\end{equation}

\subsection{Partition Functions}

The SYT labeling of supermultiplets is
convenient here as well, and it turns out that for each value
of $f\le3$ we get the precisely the
same type of multiplets as in the $OSp(6|4)$ cases considered above.
The partition functions for all of the $f \le 3$ multiplets are
\begin{equation}
\begin{aligned}
V^1 = V^1_1 &= \frac{2\sqrt{x}}{1-\sqrt{x}} \,,\\
V^2 = V^2_{1,1} &= \frac{x(3+\sqrt{x})}{1-\sqrt{x}} \,, & V^3 = V^3_{1,1,1} &= \frac{2x^\frac{3}{2}(2+\sqrt{x})}{1-\sqrt{x}} \,,\\
V^2_1 &= \frac{x(4+3\sqrt{x}+x)}{1-\sqrt{x}} \,, \quad & V^3_1 = V^3_{1,1} &= \frac{2x^\frac{3}{2}(3+3\sqrt{x}+x)}{1-\sqrt{x}} \,,\\
V^2_k &= \frac{x^\frac{k}{2}(1+\sqrt{x})^3}{1-\sqrt{x}} \,, & V^3_k = V^3_{k,1} &= \frac{2x^\frac{k+1}{2}(1+\sqrt{x})^3}{1-\sqrt{x}} \qquad k \ge 2 \,.
\end{aligned}
\end{equation}

When we move to $f=4$ however, the smaller number of superoscillator
components compared
to $OSp(6|4)$ reduces the number of possible supercovariant symmetrizations
and antisymmetrizations, leaving us with only a subset of the
types of
multiplets which appeared above.  Specifically,
only those multiplets whose lowest-weight states have $k_2 \le 2$
boxes in the second row of their super-Young tableaux are now allowed.
The partition functions of these multiplets are
\begin{equation}
\begin{aligned}
V^4 = V^4_{1,1,1,1} &= \frac{x^2(5+3\sqrt{x})}{1-\sqrt{x}} \,,\\
V^4_1 = V^4_{1,1,1} &= \frac{x^2(8+9\sqrt{x}+3x)}{1-\sqrt{x}} \,,\\
V^4_{1,1}  &= \frac{x^2(9+11\sqrt{x}+4x)}{1-\sqrt{x}} \,,\\
V^4_k = V^4_{k,1,1} &= \frac{3x^\frac{k+2}{2}(1+\sqrt{x})^3}{1-\sqrt{x}} \,,\\
V^4_{k,1} &= \frac{4x^\frac{k+2}{2}(1+\sqrt{x})^3}{1-\sqrt{x}} \,,\\
V^4_{k,2} &= \frac{x^\frac{k+2}{2}(1+\sqrt{x})^3}{1-\sqrt{x}} \,,
\end{aligned}
\end{equation}
where $k \ge 2$.
We notice that the last three partition functions are actually
proportional to each other, however the $SO(4)$
content of the corresponding representations is not the same.

\subsection{Tensor Products}

As far as the tensor products are concerned, due to the one-to-one
correspondence of multiplets for $f\le 3$ we find identical results
(when the multiplets are expressed in SYT notation)
to those
presented in~(\ref{f=2decomp1}) through~(\ref{f=3decomp2}).
For $f=4$, the existence of fewer multiplets simplifies the decompositions
slightly to
\begin{equation}
\begin{aligned}
\cv^1\tensor\ocv^1\tensor\cv^1\tensor\ocv^1
&= \sum_{j=1}^\infty \frac{1}{2} j(j+1)(\cv^4_{2j}+\ocv^4_{2j})+(j^2+j-1) \cv^4_{2j,2}+ j^2 \cv^4_{2j-1,1} \,,\\
\cv^1\tensor\cv^1\tensor\cv^1\tensor\ocv^1
&= \sum_{j=1}^\infty \frac{1}{2} j(j+1)(\cv^4_{2j-1}+\ocv^4_{2j+1})+j(j+2) \cv^4_{2j+1,2}+j(j+1)\cv^4_{2j,1} \,,\\
\cv^1\tensor\cv^1\tensor\cv^1\tensor\cv^1
&= \sum_{j=1}^\infty \frac{1}{2} j(j+1)(\cv^4_{2j-2}+\ocv^4_{2j+2})+j(j+1) \cv^4_{2j,2}+j(j+2)\cv^4_{2j+1,1} \,.
\end{aligned}
\end{equation}

It is evident that the $OSp(4|2)$ four-fold relations above can be obtained
from the respective $OSp(6|4)$ ones~(\ref{f=4decomp1})
through~(\ref{f=4decomp3})
by restricting the summation on $p$ to $p\le 1$
or equivalently to $k_2\le2$,
which is reasonable as $OSp(4|2)$ does not contain any representations
with $k_2>2$. Another way to see it is that if we tried to take the
linear combinations of superoscillators corresponding to an
$OSp(6|4)$ LWS with $k_2>2$ by using just the subset of $OSp(4|2)$
oscillators, we would get a vanishing result.

Finally the symmetrized self-conjugate $4$-fold product analogous
to the $OSp(6|4)$ result~(\ref{f=4sym_decomp}) is now
\begin{equation}
\begin{aligned}
(\cv^1\tensor\ocv^1)^2_+
&= \sum_{j=1}^\infty \lfloor {\textstyle{\frac{j+1}{2}}} \rfloor
(\lfloor{\textstyle{\frac{j+1}{2}}}\rfloor+1)(\cv^4_{2j+2}+\ocv^4_{2j+2}) \\
&\qquad\quad +(2\lfloor{\textstyle{\frac{j+1}{2}}}\rfloor^2-1)
\cv^4_{2j,2}+ \lfloor{\textstyle{\frac{j^2+1}{2}}}\rfloor \cv^4_{2j-1,1}
\\
&= \sum_{j=1}^\infty  j(j+1)(\cv^4_{4j}+\ocv^4_{4j}+\cv^4_{4j+2}+\ocv^4_{4j+2})
\\
&\qquad\quad +(2j^2-1) (\cv^4_{4j-2,2}+\cv^4_{4j,2})\\
&\qquad\quad +(2j(j-1)+1) \cv^4_{4j-3,1} + 2j^2 \cv^4_{4j-1,1} \,,
\end{aligned}
\label{osp42_symprod}
\end{equation}
where $\lfloor m \rfloor$ denotes the integer part of $m$. Again the
$OSp(4|2)$ result~(\ref{osp42_symprod})
follows from the $OSp(6|4)$
one~(\ref{f=4sym_decomp}) by simply restricting to $k_2\le 2$.

\section{A First Peek at the Two-Loop Dilatation Operator}
\label{peeksection}

Much of this paper has been rather encyclopedic in nature so
before concluding we present here an example of a
result which follows relatively easily from
information tabulated in the preceding sections.
Specifically,
we use the explicit form of the two-loop dilatation
operator~\cite{Zwiebel:2009vb,Minahan:2009te} to calculate a certain
trace $\langle D_2(x)\rangle$
of the Hamiltonian density.  This quantity enters into the
formula~\cite{Spradlin:2004pp} for the (in this case) two-loop
correction to the partition function of planar ABJM
theory on $S^2$.
It is known~\cite{Nishioka:2008gz} that, like planar SYM
on $S^3$~\cite{Sundborg:1999ue,Aharony:2003sx}, the theory has a Hagedorn
temperature $T_{\rm H}$,
which is a non-trivial function of the 't Hooft coupling $\lambda$, and
the two-loop correction to $T_{\rm H}$ at weak coupling is
determined by $\langle D_2(-1/\log T_{\rm H})\rangle$.
See~\cite{Bhattacharya:2008zy,Bhattacharya:2008bja,Hanany:2008qc,Choi:2008za}
for other work on partition functions for M2-brane theories.

\subsection{The Trace $\langle D_2(x)\rangle$ of the Hamiltonian Density}

The two-loop dilatation operator acting on a spin
chain state of length $2L$ in the ABJM theory takes the
form~\cite{Zwiebel:2009vb,Minahan:2009te}
\begin{equation}
\Delta_2  = \lambda^2 \sum_{i=1}^{2 L} (D_2)_{i,i+1,i+2}\,
\end{equation}
where the Hamiltonian density $D_2$ acts simultaneously on three adjacent
sites of the chain according to
\begin{multline}
(D_2)_{123} = \sum_{j=0}^\infty h(j)
{\cal P}^{(j)}_{12} \\
+ \sum_{j_1,j_2,j_3 = 0}^\infty
(-1)^{j_1+j_3} \left({\textstyle{\frac{1}{2}}} h(j_2-{\textstyle{\frac{1}{2}}})
+ \log 2 \right)
\left(
{\cal P}^{(j_1)}_{12}
{\cal P}^{(j_2-1/2)}_{13}
{\cal P}_{12}^{(j_3)} +
{\cal P}^{(j_1)}_{23}
{\cal P}^{(j_2-1/2)}_{13}
{\cal P}^{(j_3)}_{23}
\right),
\label{dtwo}
\end{multline}
where $h(j)$ are harmonic numbers and ${\cal P}_{ab}^{(j)}$ is
the projection operator whose image is spanned by states
with $OSp(6|4)$ spin $j$ (see~\cite{Zwiebel:2009vb} for details)
in the tensor product space of sites $a$ and $b$.

The trace we are interested in computing is
\begin{equation}
\langle D_2 (x) \rangle =
\text{Tr}_{\cv^1 \tensor \ocv^1 \tensor \cv^1}[x^\Delta D_2]\,.
\end{equation}
As a consequence of $OSp(6|4)$ symmetry
we can make use of the tensor product decomposition~(\ref{f=3decomp1}),
rewritten slightly here as
\begin{equation}
\cv^1 \tensor \ocv^1 \tensor \cv^1 = \sum_{n=0}^\infty
\lfloor {\textstyle{\frac{n+1}{2}}} \rfloor \cv(n)\,, \qquad
\cv(n) \equiv \begin{cases}
\cv^3_n & \text{for $n$ odd} \,,\\
\cv^3_{n,1} & \text{for $n$ even} \,.
\end{cases}
\label{dtwodecomp}
\end{equation}
to conclude that $D_2$ can be brought to the block-diagonal form
\begin{equation}
D_2 = \sum_{n=0}^\infty M_n \tensor {\cal P}_n
\end{equation}
where ${\cal P}_n$ is the projection operator whose image
consists of the union of the $\lfloor \frac{n+1}{2}\rfloor$ copies of
$\cv(n)$ appearing in~(\ref{dtwodecomp}) and $M_n$ is
an $\lfloor \frac{n+1}{2}\rfloor \times \lfloor \frac{n+1}{2}\rfloor$ matrix.
This form makes it clear that the desired trace can be calculated as
\begin{equation}
\langle D_2(x) \rangle = \sum_{n=0}^\infty
\text{Tr}[M_n]\, V(n)
\end{equation}
in terms of the characters which may be read off
from~(\ref{vthreecharacters})---note in particular
that $V(n) = V^3_n = V^3_{n,1}$ for $n \ge 2$. From~(\ref{dtwo}) we obtain
the values
\begin{equation}
\text{Tr}[M_n] = \left\{\begin{aligned}
2&\sum_{j=0}^{(n-2)/2} h(2j+1), && \text{$n$ even} \,,\\
2&\sum_{j=0}^{(n-1)/2} h(2j) && \text{$n$ odd}\,,
\end{aligned}
\right.
\end{equation}
which in turn lead to the result
\begin{equation}
\langle D_2(x)\rangle
=8 \sqrt{x} \frac{(1 + \sqrt{x})^2}{(1 - \sqrt{x})^6}
\left[ \sqrt{x} + x + (1 - 6 \sqrt{x} + x) \log(1 - \sqrt{x}) \right] \,.
\label{dtworesult}
\end{equation}

\subsection{The Two-Loop Hagedorn Temperature}

At zero 't Hooft coupling and infinite $N$
the partition function ${\cal Z}$
of the ABJM theory with gauge group
$U(N) \times U(N)$ on an $S^2$ of radius 1 can be expressed
(see for example~\cite{Bhattacharya:2008zy,Bhattacharya:2008bja,
Nishioka:2008gz,Choi:2008za}) as
\label{zdef}
\begin{equation}
\log {\cal Z}(x)\Big|_{\lambda = 0}
=  \log \text{Tr}[x^\Delta] \Big|_{\lambda=0}
= - \sum_{n=1}^\infty \log[1 - z(\omega^{n+1} x^n)^2]\,,
\qquad x = e^{-1/T}\,,
\end{equation}
where the trace is taken over the full Hilbert space of the ABJM
theory,
$z(x) = \frac{4 \sqrt{x}}{(1 - \sqrt{x})^2}$ is the singleton partition
function and $\omega = e^{2 \pi i}$ is a convenient bookkeeping device
which is defined
to take the value $\sqrt{\omega^m} = +1$ ($-1$) if $m$ is even (odd).
The expression~(\ref{zdef}) is valid
in the low-temperature phase $x < x_{\rm H}$, where
the Hagedorn value $x_{\rm H} = 17 - 12 \sqrt{2}$
is the smaller
solution of $z(x) = 1$.

The two-loop correction follows from
the general analysis of~\cite{Spradlin:2004pp} (see
also~\cite{Spradlin:2004sx,GomezReino:2005bq} for other
applications) and takes the form\footnote{
If the deconfinement transition of the ABJM theory at small $\lambda$
is first-order, as for example is the case~\cite{Aharony:2005bq}
for Yang-Mills theory on a small $S^3$, then ${\cal Z}(x)$ will
begin to diverge from $\text{Tr}[x^\Delta]$ for $x$ slightly below
$x_{\rm H}$, but this does not affect our calculation of the Hagedorn
temperature.}
\begin{equation}
\frac{1}{2} \frac{\partial^2}{\partial \lambda^2}
 \log \text{Tr}[x^\Delta] \Big|_{\lambda= 0}
\sim - {\log x} \sum_{n=1}^\infty
n \frac{ \langle D_2( \omega^{n+1} x^n ) \rangle z(\omega^{n+1}
x^n)}{1 - z(\omega^{n+1}
x^n)^2}\,
\label{oneloop}
\end{equation}
where $\sim$ denotes that we have omitted some additional
terms which are negligibly small as we approach the pole
in the partition function $x \to x_{\rm H}$ from below.
It follows from~(\ref{oneloop}) that the ${\cal O}(\lambda^2)$ correction
$\delta T_{\rm H}$ to the Hagedorn temperature is
\begin{equation}
\frac{\delta T_{\rm H}}{T_{\rm H}} = \frac{ \lambda^2}{\sqrt{2}}
\langle D_2(x_{\rm H}) \rangle = 2 \lambda^2 (\sqrt{2} - 1)\,,
\end{equation}
using~(\ref{dtworesult}).

\acknowledgments{
We have benefited from discussions and correspondence
with A.~Jevicki, D.~Lowe, J.~Minahan, C.~Vergu, A.~Volovich, and B.~Zwiebel.
G.~P. is grateful to the organizers of the 2008 Simons Workshop at
SUNY-Stony Brook and the Integrability in Gauge and String Theory
Workshop at the University of Utrecht for hospitality and support
during the course of this work. This work was supported in part by the US
Department of Energy under contract DE-FG02-91ER40688 and
by the US National Science Foundation grant
PHY-0638520.}

\appendix

\section{Decomposition of $OSp(6|4)$ Super-Young Tableaux}
\label{appx_submult_decomposition}

As we mentioned in section~\ref{sytsection},
the LWS of each $OSp(6|4)$ multiplet
belongs to a certain $U(2|3)$ representation, which can be
conveniently labeled by its SYT. In turn, each $U(2|3)$
representation can be decomposed into a set of $U(2)\times U(3)$
representations labeled by their respective ordinary Young tableaux.

In terms of superoscillators, this decomposition (called branching)
simply amounts to restricting their superindices
to taking either only bosonic or fermionic values in all possible
ways with distinct symmetrization and antisymmetrization properties.
This way one can immediately perform the decompositions of the first few cases,
\begin{equation}
\begin{aligned}
\syng(1)
&= \left(\yng(1),1\right)+\left(1,\yng(1)\right) \,,\\
\syng(1,1)
&= \left(\yng(1,1),1\right)+\left(\yng(1),\yng(1)\right)+\left(1,\yng(2)\right) \,,\\
\syng(2)
&= \left(\yng(2),1\right)+\left(\yng(1),\yng(1)\right)+\left(1,\yng(1,1)\right) \,,\\
\overbrace{\syng(2)\,\cdots\syng(1)}^{k\ge\,3}
&= (\overbrace{\yng(2)\,\cdots\yng(1)}^{k}\,,1)+(\overbrace{\yng(2)\,\cdots\yng(1)}^{k-1}\,,\yng(1))+(\overbrace{\yng(2)\,\cdots\yng(1)}^{k-2}\,,\yng(1,1))+(\overbrace{\yng(2)\,\cdots\yng(1)}^{k-3}\,,\yng(1,1,1)) \,,
\end{aligned}
\label{easyones}
\end{equation}
which account for all the possible lowest-weight
states which can appear for $f=1$ and $f=2$,
as one can see in section~\ref{section_reps}. For more complicated cases
one needs to use the general formulas arising from the one-to-one
correspondence between the $SU(N+M)$ and $SU(N|M)$ decompositions to
$SU(N)\times SU(M)\times U(1)$ established in~\cite{Baha Balantekin:1981bk},
or alternatively use the set of SYT decomposition rules mentioned
in~\cite{Gunaydin:1985tc}.

Here we provide a detailed list of the the additional
SYT that appear for $f=3,4$ lowest-weight states,
together with their $U(2) \times U(3)$ decompositions.
In particular, for $f=3$ we can have in addition to~(\ref{easyones})
the super-Young tableaux
\begin{equation}
\begin{aligned}
\syng(1,1,1)
&= \left(\yng(1,1),\yng(1)\right)+\left(\yng(1),\yng(2)\right)+\left(1,\yng(3)\right) \,,\\
\syng(2,1)
&= \left(\yng(2,1),1\right)+\left(\yng(2),\yng(1)\right)+\left(\yng(1,1),\yng(1)\right)+\left(\yng(1),\yng(1,1)\right)+\left(\yng(1),\yng(2)\right)+\left(1,\yng(2,1)\right) \,,\\
\syng(3,1)
&= \left(\yng(3,1),1\right)+\left(\yng(3),\yng(1)\right)+\left(\yng(2,1),\yng(1)\right)+\left(\yng(2),\yng(1,1)\right) \\
&\quad +\left(\yng(1,1),\yng(1,1)\right)+\left(\yng(2),\yng(2)\right)+\left(\yng(1),\yng(2,1)\right)+\left(\yng(1),\yng(1,1,1)\right)+\left(1,\yng(2,1,1)\right) \,,\\
\overbrace{\syng(2,1)\raise 3.2pt\hbox{\,$\cdots$\styng(1)}}^{k\ge\,4}&=(\overbrace{\yng(2,1)\raise 3.2pt\hbox{\,$\cdots$\yng(1)}}^{k}\,,1)+(\overbrace{\yng(2)\,\cdots\yng(1)}^{k}\,,\yng(1))+(\overbrace{\yng(2,1)\raise 3.2pt\hbox{\,$\cdots$\yng(1)}}^{k-1}\,,\yng(1))+(\overbrace{\yng(2)\,\cdots\yng(1)}^{k-1}\,,\yng(1,1)) \\
&\quad +(\overbrace{\yng(2)\,\cdots\yng(1)}^{k-1}\,,\yng(2))+(\overbrace{\yng(2,1)\raise 3.2pt\hbox{\,$\cdots$\yng(1)}}^{k-2}\,,\yng(1,1))+(\overbrace{\yng(2)\,\cdots\yng(1)}^{k-2}\,,\yng(2,1)) \\
&\quad +(\overbrace{\yng(2,1)\raise 3.2pt\hbox{\,$\cdots$\yng(1)}}^{k-3}\,,\yng(1,1,1))+(\overbrace{\yng(2)\,\cdots\yng(1)}^{k-2}\,,\yng(1,1,1))+(\overbrace{\yng(2)\,\cdots\yng(1)}^{k-3}\,,\yng(2,1,1)) \,,
\end{aligned}
\end{equation}
while for $f=4$ we can also have
\begin{equation}
\label{OSp64decomp}
\begin{aligned}
\syng(2,2)
&= \left(\yng(2,2),1\right)+\left(\yng(2,1),\yng(1)\right)+\left(\yng(1,1),\yng(2)\right)+\left(\yng(2),\yng(1,1)\right)+\left(\yng(1),\yng(2,1)\right)+\left(1,\yng(2,2)\right) \,,\\
\syng(3,2)
&= (\yng(3,2),1)+ (\yng(2,2),\yng(1))+ (\yng(3,1),\yng(1))+ (\yng(3),\yng(1,1))+ (\yng(2,1),\yng(1,1))+ (\yng(2,1),\yng(2)) \\
&\quad + (\yng(2),\yng(1,1,1))+ (\yng(2),\yng(2,1))+ (\yng(1,1),\yng(2,1))+ (\yng(1),\yng(2,1,1))+ (\yng(1), \yng(2,2))+ (1,\yng(2,2,1)) \,,\\
\overbrace{\syng(2,2)\raise 3.2pt\hbox{\,$\cdots$\styng(1)}}^{k\ge 4}
&= (\overbrace{\yng(2,2)\raise 3.2pt\hbox{\,$\cdots$\yng(1)}}^{k}\,,1)+(\overbrace{\yng(2,2)\raise 3.2pt\hbox{\,$\cdots$\yng(1)}}^{k-1}\,,\yng(1))+(\overbrace{\yng(2,1)\raise 3.2pt\hbox{\,$\cdots$\yng(1)}}^{k}\,,\yng(1))+(\overbrace{\yng(2)\,\cdots\yng(1)}^{k}\,,\yng(1,1)) \\
&\quad +(\overbrace{\yng(2,2)\raise 3.2pt\hbox{\,$\cdots$\yng(1)}}^{k-2}\,,\yng(1,1))+(\overbrace{\yng(2,1)\raise 3.2pt\hbox{\,$\cdots$\yng(1)}}^{k-1}\,,\yng(1,1))+(\overbrace{\yng(2,1)\raise 3.2pt\hbox{\,$\cdots$\yng(1)}}^{k-1}\,,\yng(2))+(\overbrace{\yng(2)\,\cdots\yng(1)}^{k-1}\,,\yng(2,1)) \\
&\quad +(\overbrace{\yng(2,1)\raise 3.2pt\hbox{\,$\cdots$\yng(1)}}^{k-2}\,,\yng(2,1)) +(\overbrace{\yng(2)\,\cdots\yng(1)}^{k-1}\,,\yng(1,1,1))+(\overbrace{\yng(2,1)\raise 3.2pt\hbox{\,$\cdots$\yng(1)}}^{k-2}\,,\yng(1,1,1))+(\overbrace{\yng(2,2)\raise 3.2pt\hbox{\,$\cdots$\yng(1)}}^{k-3}\,,\yng(1,1,1)) \\
&\quad +(\overbrace{\yng(2)\,\cdots\yng(1)}^{k-2}\,,\yng(2,1,1))+(\overbrace{\yng(2,1)\raise 3.2pt\hbox{\,$\cdots$\yng(1)}}^{k-3}\,,\yng(2,1,1)) +(\overbrace{\yng(2)\,\cdots\yng(1)}^{k-2}\,,\yng(2,2))+(\overbrace{\yng(2)\,\cdots\yng(1)}^{k-3}\,,\yng(2,2,1))
\end{aligned}
\end{equation}
and
\begin{equation}
\begin{aligned}
\syng(3,3)
&= \left(\yng(3,3),1\right)+\left(\yng(3,2),\yng(1)\right)+\left(\yng(2,2),\yng(2)\right)+\left(\yng(3,1),\yng(1,1)\right)+\left(\yng(2,1),\yng(2,1)\right) \\
&\quad +\left(\yng(3),\yng(1,1,1)\right)+\left(\yng(1,1),\yng(2,2)\right)+\left(\yng(2),\yng(2,1,1)\right)+\left(\yng(1),\yng(2,2,1)\right)+\left(1,\yng(2,2,2)\right) \,,\\
\syng(4,3)
&= (\yng(4,3), 1)+ (\yng(3,3), \yng(1))+ (\yng(4,2),\yng(1))+(\yng(4,1),\yng(1,1))+(\yng(3,2),\yng(1,1)) \\
&\quad +(\yng(3,2),\yng(2))+(\yng(4),\yng(1,1,1))+(\yng(3,1),\yng(1,1,1))+(\yng(2,2),\yng(2,1))+(\yng(3,1),\yng(2,1)) \\
&\quad +(\yng(3),\yng(2,1,1))+(\yng(2,1),\yng(2,1,1))+(\yng(2,1),\yng(2,2))+(\yng(2),\yng(2,2,1))+(\yng(1,1),\yng(2,2,1))+(\yng(1),\yng(2,2,2)) \,,\\
\overbrace{\syng(3,3)\raise 3.2pt\hbox{\,$\cdots$\styng(1)}}^{k\ge 5}
&= (\ayng{k}{3},1)+(\ayng{k-1}{3},\yng(1))+(\ayng{k}{2},\yng(1))+(\ayng{k}{1},\yng(1,1)) \\
&\quad+(\ayng{k-2}{3},\yng(1,1))+(\ayng{k-1}{2},\yng(1,1))+(\ayng{k-1}{2},\yng(2))+(\byng{k},\yng(1,1,1)) \\
&\quad+(\ayng{k-1}{1},\yng(1,1,1))+(\ayng{k-2}{2},\yng(1,1,1))+(\ayng{k-3}{3},\yng(1,1,1))+(\ayng{k-2}{2},\yng(2,1)) \\
&\quad+(\ayng{k-1}{1},\yng(2,1))+(\byng{k-1},\yng(2,1,1))+(\ayng{k-2}{1},\yng(2,1,1))+(\ayng{k-3}{2},\yng(2,1,1)) \\
&\quad+(\ayng{k-2}{1},  \yng(2,2))+(\byng{k-2}, \yng(2,2,1))+(\ayng{k-3}{1},\yng(2,2,1))+(\byng{k-3},\yng(2,2,2)) \,.
\end{aligned}
\label{A3}
\end{equation}
In the last lines of~(\ref{OSp64decomp}) and~(\ref{A3}) it should be understood
that a representation should be omitted from the right-hand side
if the number of boxes in the first row is less than the number in the second.
Also, we have omitted the SYT of multiplets which can be obtained
from the ones shown by conjugation, such as $\yng(2,1,1)$ and so on.

Finally, it can be shown that $U(2|3)$ representations with $k_2=3+m$
boxes in the second row of their SYT have identical decompositions as
representations with $k_2=3$ except that they have $m$
additional two-box columns in their respective $U(2)$ Young
tableaux, or alternatively in
terms of quantum numbers only the scaling dimension changes as
$\Delta\to\Delta+m$. For example, compare the decomposition of $\syng(3,3)$
above with
\begin{equation}
\begin{split}
\syng(4,4)
&= \left(\yng(4,4),1\right)+\left(\yng(4,3),\yng(1)\right)+\left(\yng(3,3),\yng(2)\right)+\left(\yng(4,2),\yng(1,1)\right)+\left(\yng(3,2),\yng(2,1)\right) \\
&\quad +\left(\yng(4,1),\yng(1,1,1)\right)+\left(\yng(2,2),\yng(2,2)\right)+\left(\yng(3,1),\yng(2,1,1)\right)+\left(\yng(2,1),\yng(2,2,1)\right)+\left(\yng(1,1),\yng(2,2,2)\right)\,,
\end{split}
\end{equation}
which evidently differs only by the addition of a single two-box column
to the $U(2)$ part of the decomposition.
With this rule one can easily obtain the decompositions of the
remaining $f=4$ SYT with $k_2 > 3$ from the results given in~(\ref{A3}).

\end{document}